\let\mypdfximage\pdfximage
\def\pdfximage{\immediate\mypdfximage}

\documentclass[usenatbib]{mn2e}
\usepackage{amsmath}
\usepackage{amssymb}
\usepackage{graphicx}
\usepackage{tensor}
\usepackage{ifthen}  
\usepackage{calc}
\usepackage{tabularx}
\usepackage{widetext}
\usepackage[hyphens]{url}
\usepackage[colorlinks,citecolor=blue]{hyperref}

\newcounter{HODDone}
\def\HOD{\ifthenelse{\equal{\arabic{HODDone}}{0}}{halo occupation distribution (HOD)\setcounter{HODDone}{1}}{HOD}}
\def\HODs{\ifthenelse{\equal{\arabic{HODDone}}{0}}{halo occupation distributions (HODs)\setcounter{HODDone}{1}}{HODs}}

\newcounter{SHMRDone}
\def\SHMR{\ifthenelse{\equal{\arabic{SHMRDone}}{0}}{stellar mass-halo mass relation (SHMR)\setcounter{SHMRDone}{1}}{SHMR}}

\newcounter{PPDDone}
\def\PPD{\ifthenelse{\equal{\arabic{PPDDone}}{0}}{posterior probability distribution (PPD)\setcounter{PPDDone}{1}}{PPD}}

\newcounter{SAMDone}
\def\SAM{\ifthenelse{\equal{\arabic{SAMDone}}{0}}{semi-analytic model (SAM)\setcounter{SAMDone}{1}}{SAM}}
\def\SAMs{\ifthenelse{\equal{\arabic{SAMDone}}{0}}{semi-analytic models (SAMs)\setcounter{SAMDone}{1}}{SAMs}}

\newcounter{MCMCDone}
\def\MCMC{\ifthenelse{\equal{\arabic{MCMCDone}}{0}}{Markov Chain Monte Carlo (MCMC)\setcounter{MCMCDone}{1}}{MCMC}}

\title{Covariances of Galaxy Stellar Mass Functions and Correlation Functions}
\author[Andrew J. Benson]{Andrew J. Benson\\
Carnegie Observatories, 813 Santa Barbara Street, Pasadena, CA 91101, USA.}

\begin{document}

\maketitle

\begin{abstract}
We compute covariance matrices for many observed estimates of the stellar mass function of galaxies from $z=0$ to $z\approx 4$, and for one estimate of the projected correlation function of galaxies split by stellar mass at $z\lesssim 0.5$. All covariance matrices include contributions due to large scale structure, the preference for galaxies to be found in groups and clusters, and for shot noise. These covariance matrices are made available for use in constraining models of galaxy formation and the galaxy-halo connection.
\end{abstract}

\begin{keywords}
galaxies: mass function, galaxies: statistics, large-scale structure of Universe
\end{keywords}

\section{Introduction}

Observational measures of population statistics of galaxies, such as the galaxy stellar mass function and correlation function, are now routinely measured with a high degree of statistical precision. These measurements are used both as quantitative constraints on theoretical models of galaxy formation \citep{bower_parameter_2010,lu_bayesian_2012,henriques_simulations_2013,mutch_constraining_2013,lu_bayesian_2014,ruiz_calibration_2015}, and in verification and validation procedures applied to mock galaxy catalogs \citep{mao_descqa:_2018}. Typically, statistical error bars\footnote{We do not address the issue of systematic errors on measurements of galaxy population statistics in this work, although they are of course extremely important.} on measurements of galaxy stellar mass functions are reported based on either the assumption that they are dominated by Poisson noise \citep[e.g.][]{baldry_galaxy_2012}, or by bootstrap procedures on mock catalogs \citep[e.g.][]{li_distribution_2009}\footnote{In some cases, estimates of systematic errors are included in the error bars reported. Since random and systematic errors affect the measurements in qualitatively different ways, it is more useful to describe them separately. For random errors, the usual approach is to give a covariance matrix (as in this work). Systematic errors are more problematic, but can often be described by a parameterized model with suitable confidence intervals for the parameters. Such systematics models can then be incorporated into Bayesian analyses of the data, with their parameters treated as hyperparameters \protect\citep{benson_building_2014,bernal_conservative_2018}.}. However, as shown by \citeauthor{smith_how_2012}~(\citeyear{smith_how_2012}; see also \citealt{benson_building_2014}), covariances between measured points in galaxy mass functions are significant in modern surveys. Ignoring these covariances when using measured mass functions to constrain galaxy formation models will lead to overly restrictive posterior distributions being derived for model parameters \citep{benson_building_2014}. Similarly, ignoring covariances can lead to false negatives in verification and validation procedures.

In this work we construct estimates for the covariance matrices of a number of different estimates of the stellar mass function of galaxies at redshift $z=0$ to $z\approx 4$, and for one estimate of the projected correlation of galaxies split by stellar mass at $z\lesssim 0.5$.  Specifically, we consider stellar mass functions of galaxies from the SDSS \citep{li_distribution_2009}, GAMA \citep{baldry_galaxy_2012}, PRIMUS \citep{moustakas_primus:_2013}, VIPERS \citep{davidzon_vimos_2013}, ULTRAVISTA \citep{muzzin_evolution_2013}, ZFOURGE \citep{tomczak_galaxy_2014}, and UKIDSS UDS \citep{caputi_stellar_2011} surveys (all available redshift bins were used from each publication),  the HI mass function of galaxies from the ALFALFA survey \citep{martin_arecibo_2010}, and projected correlation functions of galaxies as a function of stellar mass from the SDSS \citep{hearin_dark_2014}. The covariance matrices are made freely available at\ldots\footnote{Files will be made available once this paper is accepted for publication.}, and the structure of the files is detailed in Appendix~\ref{sec:fileFormat}.

\section{Methods and Results}

In this section we summarize the approach we take to computing covariance matrices, and then describe the specific details needed for each individual survey and sample being considered. We show examples of the resulting covariance matrices in two cases and discuss their qualitative features.

\subsection{Overview}

To compute covariance matrices for mass functions, we follow the same approach to this calculation as was employed in \cite{benson_building_2014} for the stellar mass function of \cite{li_distribution_2009}. That is, we use the formalism of \cite{smith_how_2012} to construct a covariance matrix containing contributions from shot noise, fluctuations due to large scale structure, and a ``halo'' term (reflecting that galaxies are correlated by virtue of the fact that they are grouped into halos). To evaluate the covariance matrix in this formalism requires knowledge of the 3D survey window function (as a function of galaxy mass), and a model of how the surveyed galaxies occupy dark matter halos. The latter we determine by fitting a parametric \HOD\ model \citep{behroozi_comprehensive_2010,leauthaud_new_2012} to the observed mass function as described in \cite{benson_building_2014}---best-fit parameters for each survey are given below. The details of survey window function construction are described below for each survey.

Computing the contribution of large scale structure to the covariance (the so-called ``cosmic variance'') requires evaluating an integral of the power spectrum over the window function of the survey. In \cite{benson_building_2014} this integral was performed by evaluating a 3-D integral in Fourier space. In Appendix~\ref{sec:lssCovariance} we derive an expression for this contribution to the covariance which involves sums over the $C_\ell$ coefficients of the spherical harmonics representation of the survey window function, and a 1-D integral over wavenumber, which is both numerically more accurate and faster to evaluate.

We have considered how well these covariances matrices describe the actual data by constructing a test-statistic of the form:
\begin{equation}
 \mathcal{T} = \Delta \mathbfss{C}^{-1} \Delta^{\rm T}
\end{equation}
where $\Delta$ is the difference between a realization of the mass function and the \HOD\ model mass function, and $\mathbfss{C}$ is the covariance matrix. We compute $\mathcal{T}_{\rm obs}$ using $\Delta=\Delta_\mathrm{obs}$ (the difference between the observed mass function and the \HOD\ model mass function), and compute a large number of realizations of $\mathcal{T}$ by generating mass functions at random from the \HOD\ model plus the covariance matrix. For many mass functions, the observed $\mathcal{T}_{\rm obs}$ lies in the low tail of the distribution of model values. This is simply due to the fact that, given a mass function with a small number of bins, our 11-parameter \HOD\ model actually over-fits the data. We do not consider this to be a significant problem, as our goal here is to simply have a reasonable description of the data. In other cases, the observations are well described  by the model. For example, in the case of the SDSS mass function, 42.6\% of model realizations exceed the observed value of the test statistic. In a handful of cases, however, we find that the observed test statistic exceeds the majority of those found from model realizations, indicating that our HOD model plus covariance matrix is an imperfect description of these datasets.

\subsection{Application to Specific Surveys}

\subsubsection{ALFALFA HI Mass Function}

For the angular mask we use the three disjoint regions defined by 07$^{\rm h}$30$^{\rm m}$ $<$ R.A. $<$ 16$^{\rm h}$30$^{\rm m}$, +04$^\circ$ $<$ decl. $<$ +16$^\circ$, and +24$^\circ$ $<$ decl. $<$ +28$^\circ$ and 22$^{\rm h}$ $<$ R.A. $<$ 03$^{\rm h}$, +14$^\circ$ $<$ decl. $<$ +16$^\circ$, and +24$^\circ$ $<$ decl. $<$ +32$^\circ$ corresponding to the sample of \cite{martin_arecibo_2010}. When the survey window function is needed we generate randomly distributed points within this angular mask and out to the survey depth. These points are used to determine which elements of a 3D grid fall within the window function.

\begin{figure}
 \includegraphics[width=85mm]{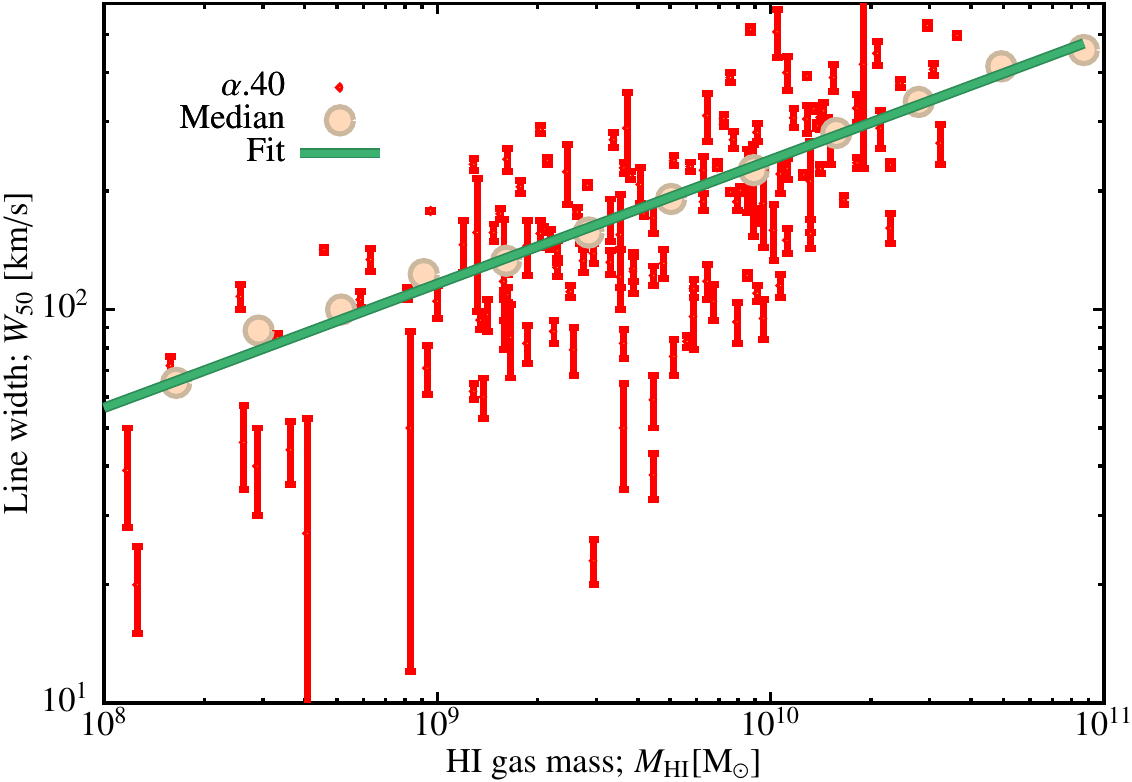}
 \caption{HI line width vs. HI mass as measured from the $\alpha.40$ survey of \protect\cite{martin_arecibo_2010}. Red points with error bars show individual measurements, while the larger circles indicate the running median of these data. The green line is a power-law fit to the running median as described in eqn.~(\protect\ref{eq:ALFALFALineWidthMassRelation}).}
 \label{fig:ALFALFALineWidthMassRelation}
\end{figure}

To estimate the depth of the \cite{martin_arecibo_2010} sample as a function of galaxy HI mass we first infer the median line width corresponding to that mass. To do so, we have fit the median line width-mass relation from the $\alpha.40$ sample with a power-law function as shown in Fig.~\ref{fig:ALFALFALineWidthMassRelation}. We find that the median line width can be approximated by
\begin{equation}
 \log_{10} (W_{\rm 50}/\hbox{km s}^{-1}) = c_0 + c_1 \log_{10}(M_{\rm HI}/{\rm M}_\odot),
 \label{eq:ALFALFALineWidthMassRelation}
\end{equation}
with $c_0=-0.770$ and $c_1=0.315$. Given the line width, the corresponding integrated flux limit, $S_{\rm int}$, for a signal-to-noise of $6.5$ is inferred using equation~(A1) of \cite{haynes_arecibo_2011}. Finally, this integrated flux limit is converted to the maximum distance at which the source could be detected using the expression given in the text of section~2.2 of \cite{martin_arecibo_2010}:
\begin{equation}
 M_{\rm HI} = 2.356\times10^5 \left(\frac{D}{\hbox{Mpc}}\right)^2 \left(\frac{S_{\rm int}}{\hbox{Jy km s}^{-1}}\right) \mathrm{M}_\odot.
\end{equation}

Priors on the parameters of the \HOD\ fit to the ALFALFA mass function of \cite{martin_arecibo_2010} are given in Table~\ref{tb:HODParameterPriors}, while the maximum likelihood values of the \HOD\ parameters are shown in Table~\ref{tb:HODMaximumLikelihood}.

\begin{table*}
 \begin{center}
 \begin{tabular}{ll@{(}r@{.}l@{,\;}r@{.}l@{)\,\,\,}l@{(}r@{.}l@{,\;}r@{.}l@{)\,\,\,}l@{(}r@{.}l@{,\;}r@{.}l@{)\,\,\,}l@{(}r@{.}l@{,\;}r@{.}l@{)\,\,\,}l@{(}r@{.}l@{,\;}r@{.}l@{)\,\,\,}l@{(}r@{.}l@{,\;}r@{.}l@{)\,\,\,}l@{(}r@{.}l@{,\;}r@{.}l@{)\,\,\,}l@{(}r@{.}l@{,\;}r@{.}l@{)\,\,\,}l@{(}r@{.}l@{,\;}r@{.}l@{)\,\,\,}l@{(}r@{.}l@{,\;}r@{.}l@{)\,\,\,}l@{(}r@{.}l@{,\;}r@{.}l@{)\,\,\,}l@{(}r@{.}l@{,\;}r@{.}l@{)\,\,\,}l@{(}r@{.}l@{,\;}r@{.}l@{)\,\,\,}l@{(}r@{.}l@{,\;}r@{.}l@{)\,\,\,}l@{(}r@{.}l@{,\;}r@{.}l@{)\,\,\,}l@{(}r@{.}l@{,\;}r@{.}l@{)\,\,\,}l@{(}r@{.}l@{,\;}r@{.}l@{)\,\,\,}l@{(}r@{.}l@{,\;}r@{.}l@{)\,\,\,}l@{(}r@{.}l@{,\;}r@{.}l@{)\,\,\,}l@{(}r@{.}l@{,\;}r@{.}l@{)\,\,\,}l@{(}r@{.}l@{,\;}r@{.}l@{)\,\,\,}l@{(}r@{.}l@{,\;}r@{.}l@{)\,\,\,}l@{(}r@{.}l@{,\;}r@{.}l@{)\,\,\,}l@{(}r@{.}l@{,\;}r@{.}l@{)\,\,\,}l@{(}r@{.}l@{,\;}r@{.}l@{)\,\,\,}l@{(}r@{.}l@{,\;}r@{.}l@{)\,\,\,}l@{(}r@{.}l@{,\;}r@{.}l@{)\,\,\,}l@{(}r@{.}l@{,\;}r@{.}l@{)\,\,\,}l@{(}r@{.}l@{,\;}r@{.}l@{)\,\,\,}l@{(}r@{.}l@{,\;}r@{.}l@{)\,\,\,}l@{(}r@{.}l@{,\;}r@{.}l@{)\,\,\,}l@{(}r@{.}l@{,\;}r@{.}l@{)\,\,\,}}
\hline
 & \multicolumn{5}{c}{\bf ALFALFA} & \multicolumn{5}{c}{\bf SDSS (Li \& White)} & \multicolumn{5}{c}{\bf SDSS (Bernardi)} & \multicolumn{5}{c}{\bf UKIDSS UDS} \\
 {\bf Parameter} & \multicolumn{5}{c}{\boldmath $z=0.00$--$0.12$} & \multicolumn{5}{c}{\boldmath $z=0.00$--$0.50$} & \multicolumn{5}{c}{\boldmath $z=0.00$--$0.50$} & \multicolumn{5}{c}{\boldmath $z=3.00$--$3.50$} \\
\hline
$\alpha_{\rm sat}$ &$N$&$0$&$98$&$0$&$000755$ &$N$&$0$&$98$&$0$&$000755$ &$N$&$0$&$98$&$0$&$000755$ &$N$&$0$&$98$&$0$&$000755$ \\
$\log_{10}(M_1/M_\odot)$ &$U$&$10$&$0$&$14$&$0$ &$U$&$12$&$1$&$12$&$6$ &$U$&$12$&$1$&$12$&$6$ &$U$&$10$&$0$&$14$&$0$ \\
$\log_{10}(M_{\star,0}/M_\odot)$ &$U$&$9$&$0$&$13$&$0$ &$U$&$10$&$5$&$10$&$8$ &$U$&$10$&$5$&$10$&$8$ &$U$&$10$&$0$&$13$&$0$ \\
$\beta$ &$U$&$0$&$0$&$1$&$0$ &$U$&$0$&$35$&$0$&$50$ &$U$&$0$&$35$&$0$&$50$ &$U$&$0$&$00$&$1$&$00$ \\
$\delta$ &$U$&$0$&$0$&$4$&$0$ &$U$&$0$&$40$&$0$&$65$ &$U$&$0$&$40$&$0$&$65$ &$U$&$0$&$0$&$4$&$0$ \\
$\gamma$ &$U$&$0$&$0$&$4$&$0$ &$U$&$0$&$7$&$1$&$9$ &$U$&$0$&$7$&$1$&$9$ &$U$&$0$&$0$&$4$&$0$ \\
$\sigma_{\log M_\star}$ &$U_{\rm ln}$&$0$&$01$&$2$&$00$ &$U_{\rm ln}$&$0$&$10$&$0$&$42$ &$U_{\rm ln}$&$0$&$10$&$0$&$42$ &$U_{\rm ln}$&$0$&$01$&$2$&$00$ \\
$B_{\rm cut}$ &$U$&$1$&$0$&$200$&$0$ &$U$&$1$&$0$&$128$&$0$ &$U$&$1$&$0$&$128$&$0$ &$U$&$1$&$0$&$128$&$0$ \\
$B_{\rm sat}$ &$U$&$0$&$0$&$15$&$0$ &$U$&$1$&$0$&$20$&$0$ &$U$&$1$&$0$&$20$&$0$ &$U$&$0$&$0$&$20$&$0$ \\
$\beta_{\rm cut}$ &$U$&$-6$&$0$&$0$&$0$ &$U$&$-2$&$0$&$0$&$0$ &$U$&$-2$&$0$&$0$&$0$ &$U$&$-6$&$0$&$0$&$0$ \\
$\beta_{\rm sat}$ &$U$&$0$&$0$&$1$&$6$ &$U$&$1$&$0$&$2$&$0$ &$U$&$1$&$0$&$2$&$0$ &$U$&$0$&$0$&$4$&$0$ \\
$\alpha_\mathrm{sb}$ &\multicolumn{5}{c}{} &\multicolumn{5}{c}{} &\multicolumn{5}{c}{} &\multicolumn{5}{c}{} \\
$\beta_\mathrm{sb}$ &\multicolumn{5}{c}{} &\multicolumn{5}{c}{} &\multicolumn{5}{c}{} &\multicolumn{5}{c}{} \\
$\gamma_\mathrm{sb}$ &\multicolumn{5}{c}{} &\multicolumn{5}{c}{} &\multicolumn{5}{c}{} &\multicolumn{5}{c}{} \\
\hline
\end{tabular}

 \begin{tabular}{ll@{(}r@{.}l@{,\;}r@{.}l@{)\,\,\,}l@{(}r@{.}l@{,\;}r@{.}l@{)\,\,\,}l@{(}r@{.}l@{,\;}r@{.}l@{)\,\,\,}l@{(}r@{.}l@{,\;}r@{.}l@{)\,\,\,}l@{(}r@{.}l@{,\;}r@{.}l@{)\,\,\,}l@{(}r@{.}l@{,\;}r@{.}l@{)\,\,\,}l@{(}r@{.}l@{,\;}r@{.}l@{)\,\,\,}l@{(}r@{.}l@{,\;}r@{.}l@{)\,\,\,}l@{(}r@{.}l@{,\;}r@{.}l@{)\,\,\,}l@{(}r@{.}l@{,\;}r@{.}l@{)\,\,\,}l@{(}r@{.}l@{,\;}r@{.}l@{)\,\,\,}l@{(}r@{.}l@{,\;}r@{.}l@{)\,\,\,}l@{(}r@{.}l@{,\;}r@{.}l@{)\,\,\,}l@{(}r@{.}l@{,\;}r@{.}l@{)\,\,\,}l@{(}r@{.}l@{,\;}r@{.}l@{)\,\,\,}l@{(}r@{.}l@{,\;}r@{.}l@{)\,\,\,}l@{(}r@{.}l@{,\;}r@{.}l@{)\,\,\,}l@{(}r@{.}l@{,\;}r@{.}l@{)\,\,\,}l@{(}r@{.}l@{,\;}r@{.}l@{)\,\,\,}l@{(}r@{.}l@{,\;}r@{.}l@{)\,\,\,}l@{(}r@{.}l@{,\;}r@{.}l@{)\,\,\,}l@{(}r@{.}l@{,\;}r@{.}l@{)\,\,\,}l@{(}r@{.}l@{,\;}r@{.}l@{)\,\,\,}l@{(}r@{.}l@{,\;}r@{.}l@{)\,\,\,}l@{(}r@{.}l@{,\;}r@{.}l@{)\,\,\,}l@{(}r@{.}l@{,\;}r@{.}l@{)\,\,\,}l@{(}r@{.}l@{,\;}r@{.}l@{)\,\,\,}l@{(}r@{.}l@{,\;}r@{.}l@{)\,\,\,}l@{(}r@{.}l@{,\;}r@{.}l@{)\,\,\,}l@{(}r@{.}l@{,\;}r@{.}l@{)\,\,\,}l@{(}r@{.}l@{,\;}r@{.}l@{)\,\,\,}l@{(}r@{.}l@{,\;}r@{.}l@{)\,\,\,}}
\hline
 & \multicolumn{10}{c}{\bf UKIDSS UDS} & \multicolumn{5}{c}{\bf GAMA} & \multicolumn{5}{c}{\bf PRIMUS} \\
 {\bf Parameter} & \multicolumn{5}{c}{\boldmath $z=3.50$--$4.25$} & \multicolumn{5}{c}{\boldmath $z=4.25$--$5.00$} & \multicolumn{5}{c}{\boldmath $z=0.00$--$0.06$} & \multicolumn{5}{c}{\boldmath $z=0.20$--$0.30$} \\
\hline
$\alpha_{\rm sat}$ &$N$&$0$&$98$&$0$&$000755$ &$N$&$0$&$98$&$0$&$000755$ &$N$&$0$&$98$&$0$&$000755$ &$N$&$0$&$98$&$0$&$000755$ \\
$\log_{10}(M_1/M_\odot)$ &$U$&$10$&$0$&$14$&$0$ &$U$&$10$&$0$&$14$&$0$ &$U$&$12$&$1$&$12$&$6$ &$U$&$10$&$0$&$14$&$0$ \\
$\log_{10}(M_{\star,0}/M_\odot)$ &$U$&$10$&$0$&$13$&$0$ &$U$&$10$&$0$&$13$&$0$ &$U$&$10$&$5$&$10$&$8$ &$U$&$10$&$0$&$13$&$0$ \\
$\beta$ &$U$&$0$&$00$&$1$&$00$ &$U$&$0$&$00$&$1$&$00$ &$U$&$0$&$35$&$0$&$50$ &$U$&$0$&$00$&$1$&$00$ \\
$\delta$ &$U$&$0$&$0$&$4$&$0$ &$U$&$0$&$0$&$4$&$0$ &$U$&$0$&$40$&$0$&$65$ &$U$&$0$&$0$&$4$&$0$ \\
$\gamma$ &$U$&$0$&$0$&$4$&$0$ &$U$&$0$&$0$&$4$&$0$ &$U$&$0$&$7$&$1$&$9$ &$U$&$0$&$0$&$4$&$0$ \\
$\sigma_{\log M_\star}$ &$U_{\rm ln}$&$0$&$01$&$2$&$00$ &$U_{\rm ln}$&$0$&$01$&$2$&$00$ &$U_{\rm ln}$&$0$&$10$&$0$&$42$ &$U_{\rm ln}$&$0$&$01$&$2$&$00$ \\
$B_{\rm cut}$ &$U$&$1$&$0$&$128$&$0$ &$U$&$1$&$0$&$128$&$0$ &$U$&$1$&$0$&$128$&$0$ &$U$&$1$&$0$&$128$&$0$ \\
$B_{\rm sat}$ &$U$&$0$&$0$&$20$&$0$ &$U$&$0$&$0$&$20$&$0$ &$U$&$1$&$0$&$20$&$0$ &$U$&$0$&$0$&$20$&$0$ \\
$\beta_{\rm cut}$ &$U$&$-6$&$0$&$0$&$0$ &$U$&$-6$&$0$&$0$&$0$ &$U$&$-2$&$0$&$0$&$0$ &$U$&$-6$&$0$&$0$&$0$ \\
$\beta_{\rm sat}$ &$U$&$0$&$0$&$4$&$0$ &$U$&$0$&$0$&$4$&$0$ &$U$&$1$&$0$&$2$&$0$ &$U$&$0$&$0$&$4$&$0$ \\
$\alpha_\mathrm{sb}$ &\multicolumn{5}{c}{} &\multicolumn{5}{c}{} &$N$&$-1$&$2$&$0$&$0225$ &\multicolumn{5}{c}{} \\
$\beta_\mathrm{sb}$ &\multicolumn{5}{c}{} &\multicolumn{5}{c}{} &$N$&$32$&$7$&$0$&$045$ &\multicolumn{5}{c}{} \\
$\gamma_\mathrm{sb}$ &\multicolumn{5}{c}{} &\multicolumn{5}{c}{} &$N$&$0$&$85$&$0$&$0025$ &\multicolumn{5}{c}{} \\
\hline
\end{tabular}

 \begin{tabular}{ll@{(}r@{.}l@{,\;}r@{.}l@{)\,\,\,}l@{(}r@{.}l@{,\;}r@{.}l@{)\,\,\,}l@{(}r@{.}l@{,\;}r@{.}l@{)\,\,\,}l@{(}r@{.}l@{,\;}r@{.}l@{)\,\,\,}l@{(}r@{.}l@{,\;}r@{.}l@{)\,\,\,}l@{(}r@{.}l@{,\;}r@{.}l@{)\,\,\,}l@{(}r@{.}l@{,\;}r@{.}l@{)\,\,\,}l@{(}r@{.}l@{,\;}r@{.}l@{)\,\,\,}l@{(}r@{.}l@{,\;}r@{.}l@{)\,\,\,}l@{(}r@{.}l@{,\;}r@{.}l@{)\,\,\,}l@{(}r@{.}l@{,\;}r@{.}l@{)\,\,\,}l@{(}r@{.}l@{,\;}r@{.}l@{)\,\,\,}l@{(}r@{.}l@{,\;}r@{.}l@{)\,\,\,}l@{(}r@{.}l@{,\;}r@{.}l@{)\,\,\,}l@{(}r@{.}l@{,\;}r@{.}l@{)\,\,\,}l@{(}r@{.}l@{,\;}r@{.}l@{)\,\,\,}l@{(}r@{.}l@{,\;}r@{.}l@{)\,\,\,}l@{(}r@{.}l@{,\;}r@{.}l@{)\,\,\,}l@{(}r@{.}l@{,\;}r@{.}l@{)\,\,\,}l@{(}r@{.}l@{,\;}r@{.}l@{)\,\,\,}l@{(}r@{.}l@{,\;}r@{.}l@{)\,\,\,}l@{(}r@{.}l@{,\;}r@{.}l@{)\,\,\,}l@{(}r@{.}l@{,\;}r@{.}l@{)\,\,\,}l@{(}r@{.}l@{,\;}r@{.}l@{)\,\,\,}l@{(}r@{.}l@{,\;}r@{.}l@{)\,\,\,}l@{(}r@{.}l@{,\;}r@{.}l@{)\,\,\,}l@{(}r@{.}l@{,\;}r@{.}l@{)\,\,\,}l@{(}r@{.}l@{,\;}r@{.}l@{)\,\,\,}l@{(}r@{.}l@{,\;}r@{.}l@{)\,\,\,}l@{(}r@{.}l@{,\;}r@{.}l@{)\,\,\,}l@{(}r@{.}l@{,\;}r@{.}l@{)\,\,\,}l@{(}r@{.}l@{,\;}r@{.}l@{)\,\,\,}}
\hline
 & \multicolumn{20}{c}{\bf PRIMUS} \\
 {\bf Parameter} & \multicolumn{5}{c}{\boldmath $z=0.30$--$0.40$} & \multicolumn{5}{c}{\boldmath $z=0.40$--$0.50$} & \multicolumn{5}{c}{\boldmath $z=0.50$--$0.65$} & \multicolumn{5}{c}{\boldmath $z=0.65$--$0.80$} \\
\hline
$\alpha_{\rm sat}$ &$N$&$0$&$98$&$0$&$000755$ &$N$&$0$&$98$&$0$&$000755$ &$N$&$0$&$98$&$0$&$000755$ &$N$&$0$&$98$&$0$&$000755$ \\
$\log_{10}(M_1/M_\odot)$ &$U$&$10$&$0$&$14$&$0$ &$U$&$10$&$0$&$14$&$0$ &$U$&$10$&$0$&$14$&$0$ &$U$&$10$&$0$&$14$&$0$ \\
$\log_{10}(M_{\star,0}/M_\odot)$ &$U$&$10$&$0$&$13$&$0$ &$U$&$10$&$0$&$13$&$0$ &$U$&$10$&$0$&$13$&$0$ &$U$&$10$&$0$&$13$&$0$ \\
$\beta$ &$U$&$0$&$00$&$1$&$00$ &$U$&$0$&$00$&$1$&$00$ &$U$&$0$&$00$&$1$&$00$ &$U$&$0$&$00$&$1$&$00$ \\
$\delta$ &$U$&$0$&$0$&$4$&$0$ &$U$&$0$&$0$&$4$&$0$ &$U$&$0$&$0$&$4$&$0$ &$U$&$0$&$0$&$4$&$0$ \\
$\gamma$ &$U$&$0$&$0$&$4$&$0$ &$U$&$0$&$0$&$4$&$0$ &$U$&$0$&$0$&$4$&$0$ &$U$&$0$&$0$&$4$&$0$ \\
$\sigma_{\log M_\star}$ &$U_{\rm ln}$&$0$&$01$&$2$&$00$ &$U_{\rm ln}$&$0$&$01$&$2$&$00$ &$U_{\rm ln}$&$0$&$01$&$2$&$00$ &$U_{\rm ln}$&$0$&$01$&$2$&$00$ \\
$B_{\rm cut}$ &$U$&$1$&$0$&$128$&$0$ &$U$&$1$&$0$&$128$&$0$ &$U$&$1$&$0$&$128$&$0$ &$U$&$1$&$0$&$128$&$0$ \\
$B_{\rm sat}$ &$U$&$0$&$0$&$20$&$0$ &$U$&$0$&$0$&$20$&$0$ &$U$&$0$&$0$&$20$&$0$ &$U$&$0$&$0$&$20$&$0$ \\
$\beta_{\rm cut}$ &$U$&$-6$&$0$&$0$&$0$ &$U$&$-6$&$0$&$0$&$0$ &$U$&$-6$&$0$&$0$&$0$ &$U$&$-6$&$0$&$0$&$0$ \\
$\beta_{\rm sat}$ &$U$&$0$&$0$&$4$&$0$ &$U$&$0$&$0$&$4$&$0$ &$U$&$0$&$0$&$4$&$0$ &$U$&$0$&$0$&$4$&$0$ \\
$\alpha_\mathrm{sb}$ &\multicolumn{5}{c}{} &\multicolumn{5}{c}{} &\multicolumn{5}{c}{} &\multicolumn{5}{c}{} \\
$\beta_\mathrm{sb}$ &\multicolumn{5}{c}{} &\multicolumn{5}{c}{} &\multicolumn{5}{c}{} &\multicolumn{5}{c}{} \\
$\gamma_\mathrm{sb}$ &\multicolumn{5}{c}{} &\multicolumn{5}{c}{} &\multicolumn{5}{c}{} &\multicolumn{5}{c}{} \\
\hline
\end{tabular}

 \end{center}
 \caption{Adopted priors for parameters of our \protect\HOD\ model. For $\alpha_{\rm sat}$, the slope of the satellite \protect\HOD\ at high masses, we adopt a prior consistent with the results of \protect\cite{kravtsov_dark_2004}. For all other parameters we adopt uniform priors spanning a wide range based on an initial estimate of the plausible ranges of the parameter values from manual tuning of the parameters. Parameters $\alpha_{\rm sat}$ through $\beta_{\rm sat}$ correspond to the \protect\HOD\ model of \protect\cite{behroozi_comprehensive_2010} and \protect\cite{leauthaud_new_2012}. Parameters $\alpha_\mathrm{sb}$, $\beta_\mathrm{sb}$, and $\gamma_\mathrm{sb}$ correspond to the surface brightness incompleteness model adopted for the GAMA survey. For the SDSS projected correlation functions of \protect\cite{hearin_dark_2014} the broad, uniform priors are chosen which span the range of the posterior distributions of parameters found by constrain the \protect\HOD\ model to match the SDSS stellar mass function of \protect\cite{li_distribution_2009}. That posterior is then applied as an additional prior over all \protect\HOD\ parameters when constraining the \protect\HOD\ model to the \protect\cite{hearin_dark_2014} correlation functions. The notation $N(\mu,s)$ indicates a normal prior with mean $\mu$ and variance, $s$, $U(a,b)$ indicates a uniform prior within the range $(a,b)$, and $U_\mathrm{ln}(a,b)$ indicates a prior which is uniform in the logarithm of the parameter within the range $(a,b)$.}
 \label{tb:HODParameterPriors}
\end{table*}

\begin{table*}
 \begin{center}
 \begin{tabular}{ll@{(}r@{.}l@{,\;}r@{.}l@{)\,\,\,}l@{(}r@{.}l@{,\;}r@{.}l@{)\,\,\,}l@{(}r@{.}l@{,\;}r@{.}l@{)\,\,\,}l@{(}r@{.}l@{,\;}r@{.}l@{)\,\,\,}l@{(}r@{.}l@{,\;}r@{.}l@{)\,\,\,}l@{(}r@{.}l@{,\;}r@{.}l@{)\,\,\,}l@{(}r@{.}l@{,\;}r@{.}l@{)\,\,\,}l@{(}r@{.}l@{,\;}r@{.}l@{)\,\,\,}l@{(}r@{.}l@{,\;}r@{.}l@{)\,\,\,}l@{(}r@{.}l@{,\;}r@{.}l@{)\,\,\,}l@{(}r@{.}l@{,\;}r@{.}l@{)\,\,\,}l@{(}r@{.}l@{,\;}r@{.}l@{)\,\,\,}l@{(}r@{.}l@{,\;}r@{.}l@{)\,\,\,}l@{(}r@{.}l@{,\;}r@{.}l@{)\,\,\,}l@{(}r@{.}l@{,\;}r@{.}l@{)\,\,\,}l@{(}r@{.}l@{,\;}r@{.}l@{)\,\,\,}l@{(}r@{.}l@{,\;}r@{.}l@{)\,\,\,}l@{(}r@{.}l@{,\;}r@{.}l@{)\,\,\,}l@{(}r@{.}l@{,\;}r@{.}l@{)\,\,\,}l@{(}r@{.}l@{,\;}r@{.}l@{)\,\,\,}l@{(}r@{.}l@{,\;}r@{.}l@{)\,\,\,}l@{(}r@{.}l@{,\;}r@{.}l@{)\,\,\,}l@{(}r@{.}l@{,\;}r@{.}l@{)\,\,\,}l@{(}r@{.}l@{,\;}r@{.}l@{)\,\,\,}l@{(}r@{.}l@{,\;}r@{.}l@{)\,\,\,}l@{(}r@{.}l@{,\;}r@{.}l@{)\,\,\,}l@{(}r@{.}l@{,\;}r@{.}l@{)\,\,\,}l@{(}r@{.}l@{,\;}r@{.}l@{)\,\,\,}l@{(}r@{.}l@{,\;}r@{.}l@{)\,\,\,}l@{(}r@{.}l@{,\;}r@{.}l@{)\,\,\,}l@{(}r@{.}l@{,\;}r@{.}l@{)\,\,\,}l@{(}r@{.}l@{,\;}r@{.}l@{)\,\,\,}}
\hline
 & \multicolumn{5}{c}{\bf PRIMUS} & \multicolumn{15}{c}{\bf VIPERS} \\
 {\bf Parameter} & \multicolumn{5}{c}{\boldmath $z=0.80$--$1.00$} & \multicolumn{5}{c}{\boldmath $z=0.50$--$0.60$} & \multicolumn{5}{c}{\boldmath $z=0.60$--$0.80$} & \multicolumn{5}{c}{\boldmath $z=0.80$--$1.00$} \\
\hline
$\alpha_{\rm sat}$ &$N$&$0$&$98$&$0$&$000755$ &$N$&$0$&$98$&$0$&$000755$ &$N$&$0$&$98$&$0$&$000755$ &$N$&$0$&$98$&$0$&$000755$ \\
$\log_{10}(M_1/M_\odot)$ &$U$&$10$&$0$&$14$&$0$ &$U$&$10$&$0$&$14$&$0$ &$U$&$10$&$0$&$14$&$0$ &$U$&$10$&$0$&$14$&$0$ \\
$\log_{10}(M_{\star,0}/M_\odot)$ &$U$&$10$&$0$&$13$&$0$ &$U$&$10$&$0$&$13$&$0$ &$U$&$10$&$0$&$13$&$0$ &$U$&$10$&$0$&$13$&$0$ \\
$\beta$ &$U$&$0$&$00$&$1$&$00$ &$U$&$0$&$00$&$1$&$00$ &$U$&$0$&$00$&$1$&$00$ &$U$&$0$&$00$&$1$&$00$ \\
$\delta$ &$U$&$0$&$0$&$4$&$0$ &$U$&$0$&$0$&$4$&$0$ &$U$&$0$&$0$&$4$&$0$ &$U$&$0$&$0$&$4$&$0$ \\
$\gamma$ &$U$&$0$&$0$&$4$&$0$ &$U$&$0$&$0$&$4$&$0$ &$U$&$0$&$0$&$4$&$0$ &$U$&$0$&$0$&$4$&$0$ \\
$\sigma_{\log M_\star}$ &$U_{\rm ln}$&$0$&$01$&$2$&$00$ &$U_{\rm ln}$&$0$&$01$&$2$&$00$ &$U_{\rm ln}$&$0$&$01$&$2$&$00$ &$U_{\rm ln}$&$0$&$01$&$2$&$00$ \\
$B_{\rm cut}$ &$U$&$1$&$0$&$128$&$0$ &$U$&$1$&$0$&$128$&$0$ &$U$&$1$&$0$&$128$&$0$ &$U$&$1$&$0$&$128$&$0$ \\
$B_{\rm sat}$ &$U$&$0$&$0$&$20$&$0$ &$U$&$0$&$0$&$20$&$0$ &$U$&$0$&$0$&$20$&$0$ &$U$&$0$&$0$&$20$&$0$ \\
$\beta_{\rm cut}$ &$U$&$-6$&$0$&$0$&$0$ &$U$&$-6$&$0$&$0$&$0$ &$U$&$-6$&$0$&$0$&$0$ &$U$&$-6$&$0$&$0$&$0$ \\
$\beta_{\rm sat}$ &$U$&$0$&$0$&$4$&$0$ &$U$&$0$&$0$&$4$&$0$ &$U$&$0$&$0$&$4$&$0$ &$U$&$0$&$0$&$4$&$0$ \\
$\alpha_\mathrm{sb}$ &\multicolumn{5}{c}{} &\multicolumn{5}{c}{} &\multicolumn{5}{c}{} &\multicolumn{5}{c}{} \\
$\beta_\mathrm{sb}$ &\multicolumn{5}{c}{} &\multicolumn{5}{c}{} &\multicolumn{5}{c}{} &\multicolumn{5}{c}{} \\
$\gamma_\mathrm{sb}$ &\multicolumn{5}{c}{} &\multicolumn{5}{c}{} &\multicolumn{5}{c}{} &\multicolumn{5}{c}{} \\
\hline
\end{tabular}

 \begin{tabular}{ll@{(}r@{.}l@{,\;}r@{.}l@{)\,\,\,}l@{(}r@{.}l@{,\;}r@{.}l@{)\,\,\,}l@{(}r@{.}l@{,\;}r@{.}l@{)\,\,\,}l@{(}r@{.}l@{,\;}r@{.}l@{)\,\,\,}l@{(}r@{.}l@{,\;}r@{.}l@{)\,\,\,}l@{(}r@{.}l@{,\;}r@{.}l@{)\,\,\,}l@{(}r@{.}l@{,\;}r@{.}l@{)\,\,\,}l@{(}r@{.}l@{,\;}r@{.}l@{)\,\,\,}l@{(}r@{.}l@{,\;}r@{.}l@{)\,\,\,}l@{(}r@{.}l@{,\;}r@{.}l@{)\,\,\,}l@{(}r@{.}l@{,\;}r@{.}l@{)\,\,\,}l@{(}r@{.}l@{,\;}r@{.}l@{)\,\,\,}l@{(}r@{.}l@{,\;}r@{.}l@{)\,\,\,}l@{(}r@{.}l@{,\;}r@{.}l@{)\,\,\,}l@{(}r@{.}l@{,\;}r@{.}l@{)\,\,\,}l@{(}r@{.}l@{,\;}r@{.}l@{)\,\,\,}l@{(}r@{.}l@{,\;}r@{.}l@{)\,\,\,}l@{(}r@{.}l@{,\;}r@{.}l@{)\,\,\,}l@{(}r@{.}l@{,\;}r@{.}l@{)\,\,\,}l@{(}r@{.}l@{,\;}r@{.}l@{)\,\,\,}l@{(}r@{.}l@{,\;}r@{.}l@{)\,\,\,}l@{(}r@{.}l@{,\;}r@{.}l@{)\,\,\,}l@{(}r@{.}l@{,\;}r@{.}l@{)\,\,\,}l@{(}r@{.}l@{,\;}r@{.}l@{)\,\,\,}l@{(}r@{.}l@{,\;}r@{.}l@{)\,\,\,}l@{(}r@{.}l@{,\;}r@{.}l@{)\,\,\,}l@{(}r@{.}l@{,\;}r@{.}l@{)\,\,\,}l@{(}r@{.}l@{,\;}r@{.}l@{)\,\,\,}l@{(}r@{.}l@{,\;}r@{.}l@{)\,\,\,}l@{(}r@{.}l@{,\;}r@{.}l@{)\,\,\,}l@{(}r@{.}l@{,\;}r@{.}l@{)\,\,\,}l@{(}r@{.}l@{,\;}r@{.}l@{)\,\,\,}}
\hline
 & \multicolumn{20}{c}{\bf ZFOURGE} \\
 {\bf Parameter} & \multicolumn{5}{c}{\boldmath $z=0.20$--$0.50$} & \multicolumn{5}{c}{\boldmath $z=0.50$--$0.75$} & \multicolumn{5}{c}{\boldmath $z=0.75$--$1.00$} & \multicolumn{5}{c}{\boldmath $z=1.00$--$1.25$} \\
\hline
$\alpha_{\rm sat}$ &$N$&$0$&$98$&$0$&$000755$ &$N$&$0$&$98$&$0$&$000755$ &$N$&$0$&$98$&$0$&$000755$ &$N$&$0$&$98$&$0$&$000755$ \\
$\log_{10}(M_1/M_\odot)$ &$U$&$10$&$0$&$14$&$0$ &$U$&$10$&$0$&$14$&$0$ &$U$&$10$&$0$&$14$&$0$ &$U$&$10$&$0$&$14$&$0$ \\
$\log_{10}(M_{\star,0}/M_\odot)$ &$U$&$10$&$0$&$13$&$0$ &$U$&$10$&$0$&$13$&$0$ &$U$&$10$&$0$&$13$&$0$ &$U$&$10$&$0$&$13$&$0$ \\
$\beta$ &$U$&$0$&$00$&$1$&$00$ &$U$&$0$&$00$&$1$&$00$ &$U$&$0$&$00$&$1$&$00$ &$U$&$0$&$00$&$1$&$00$ \\
$\delta$ &$U$&$0$&$0$&$4$&$0$ &$U$&$0$&$0$&$4$&$0$ &$U$&$0$&$0$&$4$&$0$ &$U$&$0$&$0$&$4$&$0$ \\
$\gamma$ &$U$&$0$&$0$&$4$&$0$ &$U$&$0$&$0$&$4$&$0$ &$U$&$0$&$0$&$4$&$0$ &$U$&$0$&$0$&$4$&$0$ \\
$\sigma_{\log M_\star}$ &$U_{\rm ln}$&$0$&$01$&$2$&$00$ &$U_{\rm ln}$&$0$&$01$&$2$&$00$ &$U_{\rm ln}$&$0$&$01$&$2$&$00$ &$U_{\rm ln}$&$0$&$01$&$2$&$00$ \\
$B_{\rm cut}$ &$U$&$1$&$0$&$128$&$0$ &$U$&$1$&$0$&$128$&$0$ &$U$&$1$&$0$&$128$&$0$ &$U$&$1$&$0$&$128$&$0$ \\
$B_{\rm sat}$ &$U$&$0$&$0$&$20$&$0$ &$U$&$0$&$0$&$20$&$0$ &$U$&$0$&$0$&$20$&$0$ &$U$&$0$&$0$&$20$&$0$ \\
$\beta_{\rm cut}$ &$U$&$-6$&$0$&$0$&$0$ &$U$&$-6$&$0$&$0$&$0$ &$U$&$-6$&$0$&$0$&$0$ &$U$&$-6$&$0$&$0$&$0$ \\
$\beta_{\rm sat}$ &$U$&$0$&$0$&$4$&$0$ &$U$&$0$&$0$&$4$&$0$ &$U$&$0$&$0$&$4$&$0$ &$U$&$0$&$0$&$4$&$0$ \\
$\alpha_\mathrm{sb}$ &\multicolumn{5}{c}{} &\multicolumn{5}{c}{} &\multicolumn{5}{c}{} &\multicolumn{5}{c}{} \\
$\beta_\mathrm{sb}$ &\multicolumn{5}{c}{} &\multicolumn{5}{c}{} &\multicolumn{5}{c}{} &\multicolumn{5}{c}{} \\
$\gamma_\mathrm{sb}$ &\multicolumn{5}{c}{} &\multicolumn{5}{c}{} &\multicolumn{5}{c}{} &\multicolumn{5}{c}{} \\
\hline
\end{tabular}

 \begin{tabular}{ll@{(}r@{.}l@{,\;}r@{.}l@{)\,\,\,}l@{(}r@{.}l@{,\;}r@{.}l@{)\,\,\,}l@{(}r@{.}l@{,\;}r@{.}l@{)\,\,\,}l@{(}r@{.}l@{,\;}r@{.}l@{)\,\,\,}l@{(}r@{.}l@{,\;}r@{.}l@{)\,\,\,}l@{(}r@{.}l@{,\;}r@{.}l@{)\,\,\,}l@{(}r@{.}l@{,\;}r@{.}l@{)\,\,\,}l@{(}r@{.}l@{,\;}r@{.}l@{)\,\,\,}l@{(}r@{.}l@{,\;}r@{.}l@{)\,\,\,}l@{(}r@{.}l@{,\;}r@{.}l@{)\,\,\,}l@{(}r@{.}l@{,\;}r@{.}l@{)\,\,\,}l@{(}r@{.}l@{,\;}r@{.}l@{)\,\,\,}l@{(}r@{.}l@{,\;}r@{.}l@{)\,\,\,}l@{(}r@{.}l@{,\;}r@{.}l@{)\,\,\,}l@{(}r@{.}l@{,\;}r@{.}l@{)\,\,\,}l@{(}r@{.}l@{,\;}r@{.}l@{)\,\,\,}l@{(}r@{.}l@{,\;}r@{.}l@{)\,\,\,}l@{(}r@{.}l@{,\;}r@{.}l@{)\,\,\,}l@{(}r@{.}l@{,\;}r@{.}l@{)\,\,\,}l@{(}r@{.}l@{,\;}r@{.}l@{)\,\,\,}l@{(}r@{.}l@{,\;}r@{.}l@{)\,\,\,}l@{(}r@{.}l@{,\;}r@{.}l@{)\,\,\,}l@{(}r@{.}l@{,\;}r@{.}l@{)\,\,\,}l@{(}r@{.}l@{,\;}r@{.}l@{)\,\,\,}l@{(}r@{.}l@{,\;}r@{.}l@{)\,\,\,}l@{(}r@{.}l@{,\;}r@{.}l@{)\,\,\,}l@{(}r@{.}l@{,\;}r@{.}l@{)\,\,\,}l@{(}r@{.}l@{,\;}r@{.}l@{)\,\,\,}l@{(}r@{.}l@{,\;}r@{.}l@{)\,\,\,}l@{(}r@{.}l@{,\;}r@{.}l@{)\,\,\,}l@{(}r@{.}l@{,\;}r@{.}l@{)\,\,\,}l@{(}r@{.}l@{,\;}r@{.}l@{)\,\,\,}}
\hline
 & \multicolumn{20}{c}{\bf ZFOURGE} \\
 {\bf Parameter} & \multicolumn{5}{c}{\boldmath $z=1.25$--$1.50$} & \multicolumn{5}{c}{\boldmath $z=1.50$--$2.00$} & \multicolumn{5}{c}{\boldmath $z=2.00$--$2.50$} & \multicolumn{5}{c}{\boldmath $z=2.50$--$3.00$} \\
\hline
$\alpha_{\rm sat}$ &$N$&$0$&$98$&$0$&$000755$ &$N$&$0$&$98$&$0$&$000755$ &$N$&$0$&$98$&$0$&$000755$ &$N$&$0$&$98$&$0$&$000755$ \\
$\log_{10}(M_1/M_\odot)$ &$U$&$10$&$0$&$14$&$0$ &$U$&$10$&$0$&$14$&$0$ &$U$&$10$&$0$&$14$&$0$ &$U$&$10$&$0$&$14$&$0$ \\
$\log_{10}(M_{\star,0}/M_\odot)$ &$U$&$10$&$0$&$13$&$0$ &$U$&$10$&$0$&$13$&$0$ &$U$&$10$&$0$&$13$&$0$ &$U$&$10$&$0$&$13$&$0$ \\
$\beta$ &$U$&$0$&$00$&$1$&$00$ &$U$&$0$&$00$&$1$&$00$ &$U$&$0$&$00$&$1$&$00$ &$U$&$0$&$00$&$1$&$00$ \\
$\delta$ &$U$&$0$&$0$&$4$&$0$ &$U$&$0$&$0$&$4$&$0$ &$U$&$0$&$0$&$4$&$0$ &$U$&$0$&$0$&$4$&$0$ \\
$\gamma$ &$U$&$0$&$0$&$4$&$0$ &$U$&$0$&$0$&$4$&$0$ &$U$&$0$&$0$&$4$&$0$ &$U$&$0$&$0$&$4$&$0$ \\
$\sigma_{\log M_\star}$ &$U_{\rm ln}$&$0$&$01$&$2$&$00$ &$U_{\rm ln}$&$0$&$01$&$2$&$00$ &$U_{\rm ln}$&$0$&$01$&$2$&$00$ &$U_{\rm ln}$&$0$&$01$&$2$&$00$ \\
$B_{\rm cut}$ &$U$&$1$&$0$&$128$&$0$ &$U$&$1$&$0$&$128$&$0$ &$U$&$1$&$0$&$128$&$0$ &$U$&$1$&$0$&$128$&$0$ \\
$B_{\rm sat}$ &$U$&$0$&$0$&$20$&$0$ &$U$&$0$&$0$&$20$&$0$ &$U$&$0$&$0$&$20$&$0$ &$U$&$0$&$0$&$20$&$0$ \\
$\beta_{\rm cut}$ &$U$&$-6$&$0$&$0$&$0$ &$U$&$-6$&$0$&$0$&$0$ &$U$&$-6$&$0$&$0$&$0$ &$U$&$-6$&$0$&$0$&$0$ \\
$\beta_{\rm sat}$ &$U$&$0$&$0$&$4$&$0$ &$U$&$0$&$0$&$4$&$0$ &$U$&$0$&$0$&$4$&$0$ &$U$&$0$&$0$&$4$&$0$ \\
$\alpha_\mathrm{sb}$ &\multicolumn{5}{c}{} &\multicolumn{5}{c}{} &\multicolumn{5}{c}{} &\multicolumn{5}{c}{} \\
$\beta_\mathrm{sb}$ &\multicolumn{5}{c}{} &\multicolumn{5}{c}{} &\multicolumn{5}{c}{} &\multicolumn{5}{c}{} \\
$\gamma_\mathrm{sb}$ &\multicolumn{5}{c}{} &\multicolumn{5}{c}{} &\multicolumn{5}{c}{} &\multicolumn{5}{c}{} \\
\hline
\end{tabular}

 \end{center}
 \addtocounter{table}{-1}
 \caption{\emph{(cont.)}}
\end{table*}

\begin{table*}
 \begin{center}
 \begin{tabular}{ll@{(}r@{.}l@{,\;}r@{.}l@{)\,\,\,}l@{(}r@{.}l@{,\;}r@{.}l@{)\,\,\,}l@{(}r@{.}l@{,\;}r@{.}l@{)\,\,\,}l@{(}r@{.}l@{,\;}r@{.}l@{)\,\,\,}l@{(}r@{.}l@{,\;}r@{.}l@{)\,\,\,}l@{(}r@{.}l@{,\;}r@{.}l@{)\,\,\,}l@{(}r@{.}l@{,\;}r@{.}l@{)\,\,\,}l@{(}r@{.}l@{,\;}r@{.}l@{)\,\,\,}l@{(}r@{.}l@{,\;}r@{.}l@{)\,\,\,}l@{(}r@{.}l@{,\;}r@{.}l@{)\,\,\,}l@{(}r@{.}l@{,\;}r@{.}l@{)\,\,\,}l@{(}r@{.}l@{,\;}r@{.}l@{)\,\,\,}l@{(}r@{.}l@{,\;}r@{.}l@{)\,\,\,}l@{(}r@{.}l@{,\;}r@{.}l@{)\,\,\,}l@{(}r@{.}l@{,\;}r@{.}l@{)\,\,\,}l@{(}r@{.}l@{,\;}r@{.}l@{)\,\,\,}l@{(}r@{.}l@{,\;}r@{.}l@{)\,\,\,}l@{(}r@{.}l@{,\;}r@{.}l@{)\,\,\,}l@{(}r@{.}l@{,\;}r@{.}l@{)\,\,\,}l@{(}r@{.}l@{,\;}r@{.}l@{)\,\,\,}l@{(}r@{.}l@{,\;}r@{.}l@{)\,\,\,}l@{(}r@{.}l@{,\;}r@{.}l@{)\,\,\,}l@{(}r@{.}l@{,\;}r@{.}l@{)\,\,\,}l@{(}r@{.}l@{,\;}r@{.}l@{)\,\,\,}l@{(}r@{.}l@{,\;}r@{.}l@{)\,\,\,}l@{(}r@{.}l@{,\;}r@{.}l@{)\,\,\,}l@{(}r@{.}l@{,\;}r@{.}l@{)\,\,\,}l@{(}r@{.}l@{,\;}r@{.}l@{)\,\,\,}l@{(}r@{.}l@{,\;}r@{.}l@{)\,\,\,}l@{(}r@{.}l@{,\;}r@{.}l@{)\,\,\,}l@{(}r@{.}l@{,\;}r@{.}l@{)\,\,\,}l@{(}r@{.}l@{,\;}r@{.}l@{)\,\,\,}}
\hline
 & \multicolumn{20}{c}{\bf ULTRAVISTA} \\
 {\bf Parameter} & \multicolumn{5}{c}{\boldmath $z=0.20$--$0.50$} & \multicolumn{5}{c}{\boldmath $z=0.50$--$1.00$} & \multicolumn{5}{c}{\boldmath $z=1.00$--$1.50$} & \multicolumn{5}{c}{\boldmath $z=1.50$--$2.00$} \\
\hline
$\alpha_{\rm sat}$ &$N$&$0$&$98$&$0$&$000755$ &$N$&$0$&$98$&$0$&$000755$ &$N$&$0$&$98$&$0$&$000755$ &$N$&$0$&$98$&$0$&$000755$ \\
$\log_{10}(M_1/M_\odot)$ &$U$&$10$&$0$&$14$&$0$ &$U$&$10$&$0$&$14$&$0$ &$U$&$10$&$0$&$14$&$0$ &$U$&$10$&$0$&$14$&$0$ \\
$\log_{10}(M_{\star,0}/M_\odot)$ &$U$&$10$&$0$&$13$&$0$ &$U$&$10$&$0$&$13$&$0$ &$U$&$10$&$0$&$13$&$0$ &$U$&$10$&$0$&$13$&$0$ \\
$\beta$ &$U$&$0$&$00$&$1$&$00$ &$U$&$0$&$00$&$1$&$00$ &$U$&$0$&$00$&$1$&$00$ &$U$&$0$&$00$&$1$&$00$ \\
$\delta$ &$U$&$0$&$0$&$4$&$0$ &$U$&$0$&$0$&$6$&$0$ &$U$&$0$&$0$&$4$&$0$ &$U$&$0$&$0$&$6$&$0$ \\
$\gamma$ &$U$&$0$&$0$&$4$&$0$ &$U$&$0$&$0$&$6$&$0$ &$U$&$0$&$0$&$4$&$0$ &$U$&$0$&$0$&$6$&$0$ \\
$\sigma_{\log M_\star}$ &$U_{\rm ln}$&$0$&$01$&$2$&$00$ &$U_{\rm ln}$&$0$&$01$&$2$&$00$ &$U_{\rm ln}$&$0$&$01$&$2$&$00$ &$U_{\rm ln}$&$0$&$01$&$2$&$00$ \\
$B_{\rm cut}$ &$U$&$1$&$0$&$128$&$0$ &$U$&$1$&$0$&$128$&$0$ &$U$&$1$&$0$&$128$&$0$ &$U$&$1$&$0$&$128$&$0$ \\
$B_{\rm sat}$ &$U$&$0$&$0$&$20$&$0$ &$U$&$0$&$0$&$20$&$0$ &$U$&$0$&$0$&$20$&$0$ &$U$&$0$&$0$&$20$&$0$ \\
$\beta_{\rm cut}$ &$U$&$-6$&$0$&$0$&$0$ &$U$&$-10$&$0$&$0$&$0$ &$U$&$-6$&$0$&$0$&$0$ &$U$&$-10$&$0$&$0$&$0$ \\
$\beta_{\rm sat}$ &$U$&$0$&$0$&$4$&$0$ &$U$&$0$&$0$&$4$&$0$ &$U$&$0$&$0$&$4$&$0$ &$U$&$0$&$0$&$4$&$0$ \\
$\alpha_\mathrm{sb}$ &\multicolumn{5}{c}{} &\multicolumn{5}{c}{} &\multicolumn{5}{c}{} &\multicolumn{5}{c}{} \\
$\beta_\mathrm{sb}$ &\multicolumn{5}{c}{} &\multicolumn{5}{c}{} &\multicolumn{5}{c}{} &\multicolumn{5}{c}{} \\
$\gamma_\mathrm{sb}$ &\multicolumn{5}{c}{} &\multicolumn{5}{c}{} &\multicolumn{5}{c}{} &\multicolumn{5}{c}{} \\
\hline
\end{tabular}

 \begin{tabular}{ll@{(}r@{.}l@{,\;}r@{.}l@{)\,\,\,}l@{(}r@{.}l@{,\;}r@{.}l@{)\,\,\,}l@{(}r@{.}l@{,\;}r@{.}l@{)\,\,\,}l@{(}r@{.}l@{,\;}r@{.}l@{)\,\,\,}l@{(}r@{.}l@{,\;}r@{.}l@{)\,\,\,}l@{(}r@{.}l@{,\;}r@{.}l@{)\,\,\,}l@{(}r@{.}l@{,\;}r@{.}l@{)\,\,\,}l@{(}r@{.}l@{,\;}r@{.}l@{)\,\,\,}l@{(}r@{.}l@{,\;}r@{.}l@{)\,\,\,}l@{(}r@{.}l@{,\;}r@{.}l@{)\,\,\,}l@{(}r@{.}l@{,\;}r@{.}l@{)\,\,\,}l@{(}r@{.}l@{,\;}r@{.}l@{)\,\,\,}l@{(}r@{.}l@{,\;}r@{.}l@{)\,\,\,}l@{(}r@{.}l@{,\;}r@{.}l@{)\,\,\,}l@{(}r@{.}l@{,\;}r@{.}l@{)\,\,\,}l@{(}r@{.}l@{,\;}r@{.}l@{)\,\,\,}l@{(}r@{.}l@{,\;}r@{.}l@{)\,\,\,}l@{(}r@{.}l@{,\;}r@{.}l@{)\,\,\,}l@{(}r@{.}l@{,\;}r@{.}l@{)\,\,\,}l@{(}r@{.}l@{,\;}r@{.}l@{)\,\,\,}l@{(}r@{.}l@{,\;}r@{.}l@{)\,\,\,}l@{(}r@{.}l@{,\;}r@{.}l@{)\,\,\,}l@{(}r@{.}l@{,\;}r@{.}l@{)\,\,\,}l@{(}r@{.}l@{,\;}r@{.}l@{)\,\,\,}l@{(}r@{.}l@{,\;}r@{.}l@{)\,\,\,}l@{(}r@{.}l@{,\;}r@{.}l@{)\,\,\,}l@{(}r@{.}l@{,\;}r@{.}l@{)\,\,\,}l@{(}r@{.}l@{,\;}r@{.}l@{)\,\,\,}l@{(}r@{.}l@{,\;}r@{.}l@{)\,\,\,}l@{(}r@{.}l@{,\;}r@{.}l@{)\,\,\,}l@{(}r@{.}l@{,\;}r@{.}l@{)\,\,\,}l@{(}r@{.}l@{,\;}r@{.}l@{)\,\,\,}}
\hline
 & \multicolumn{15}{c}{\bf ULTRAVISTA} & \multicolumn{5}{c}{\bf SDSS (Hearin et al.)} \\
 {\bf Parameter} & \multicolumn{5}{c}{\boldmath $z=2.00$--$2.50$} & \multicolumn{5}{c}{\boldmath $z=2.50$--$3.00$} & \multicolumn{5}{c}{\boldmath $z=3.00$--$4.00$} & \multicolumn{5}{c}{\boldmath $z=0.00$--$0.50$} \\
\hline
$\alpha_{\rm sat}$ &$N$&$0$&$98$&$0$&$000755$ &$N$&$0$&$98$&$0$&$000755$ &$N$&$0$&$98$&$0$&$000755$ &$U$&$0$&$5$&$2$&$0$ \\
$\log_{10}(M_1/M_\odot)$ &$U$&$10$&$0$&$14$&$0$ &$U$&$10$&$0$&$14$&$0$ &$U$&$10$&$0$&$14$&$0$ &$U$&$11$&$0$&$13$&$0$ \\
$\log_{10}(M_{\star,0}/M_\odot)$ &$U$&$10$&$0$&$13$&$0$ &$U$&$10$&$0$&$13$&$0$ &$U$&$10$&$0$&$13$&$0$ &$U$&$9$&$5$&$12$&$0$ \\
$\beta$ &$U$&$0$&$00$&$1$&$00$ &$U$&$0$&$00$&$1$&$00$ &$U$&$0$&$00$&$1$&$00$ &$U$&$0$&$00$&$1$&$00$ \\
$\delta$ &$U$&$0$&$0$&$4$&$0$ &$U$&$0$&$0$&$6$&$0$ &$U$&$0$&$0$&$4$&$0$ &$U$&$0$&$0$&$2$&$0$ \\
$\gamma$ &$U$&$0$&$0$&$4$&$0$ &$U$&$0$&$0$&$6$&$0$ &$U$&$0$&$0$&$4$&$0$ &$U$&$0$&$0$&$3$&$0$ \\
$\sigma_{\log M_\star}$ &$U_{\rm ln}$&$0$&$01$&$2$&$00$ &$U_{\rm ln}$&$0$&$01$&$2$&$00$ &$U_{\rm ln}$&$0$&$01$&$2$&$00$ &$U_{\rm ln}$&$0$&$10$&$0$&$42$ \\
$B_{\rm cut}$ &$U$&$1$&$0$&$128$&$0$ &$U$&$1$&$0$&$128$&$0$ &$U$&$1$&$0$&$128$&$0$ &$U$&$90$&$0$&$130$&$0$ \\
$B_{\rm sat}$ &$U$&$0$&$0$&$20$&$0$ &$U$&$0$&$0$&$20$&$0$ &$U$&$0$&$0$&$20$&$0$ &$U$&$5$&$0$&$30$&$0$ \\
$\beta_{\rm cut}$ &$U$&$-6$&$0$&$0$&$0$ &$U$&$-10$&$0$&$0$&$0$ &$U$&$-6$&$0$&$0$&$0$ &$U$&$-3$&$0$&$-1$&$0$ \\
$\beta_{\rm sat}$ &$U$&$0$&$0$&$4$&$0$ &$U$&$0$&$0$&$4$&$0$ &$U$&$0$&$0$&$4$&$0$ &$U$&$1$&$0$&$2$&$0$ \\
$\alpha_\mathrm{sb}$ &\multicolumn{5}{c}{} &\multicolumn{5}{c}{} &\multicolumn{5}{c}{} &\multicolumn{5}{c}{} \\
$\beta_\mathrm{sb}$ &\multicolumn{5}{c}{} &\multicolumn{5}{c}{} &\multicolumn{5}{c}{} &\multicolumn{5}{c}{} \\
$\gamma_\mathrm{sb}$ &\multicolumn{5}{c}{} &\multicolumn{5}{c}{} &\multicolumn{5}{c}{} &\multicolumn{5}{c}{} \\
\hline
\end{tabular}

 \end{center}
 \addtocounter{table}{-1}
 \caption{\emph{(cont.)}}
\end{table*}

\begin{table*}
 \begin{center}
 \begin{tabular}{lr@{.}lr@{.}lr@{.}lr@{.}lr@{.}l}
\hline
 & \multicolumn{2}{c}{\textbf{ALFALFA}} & \multicolumn{2}{c}{\textbf{SDSS (Li \& White)}} & \multicolumn{2}{c}{\textbf{SDSS (Bernardi et al.)}} & \multicolumn{2}{c}{\textbf{UKIDSS UDS}} & \multicolumn{2}{c}{\textbf{UKIDSS UDS}} \\
 \textbf{Parameter} & \multicolumn{2}{c}{\boldmath $z=0.00$--$0.12$} & \multicolumn{2}{c}{\boldmath $z=0.00$--$0.50$} & \multicolumn{2}{c}{\boldmath $z=0.00$--$0.50$} & \multicolumn{2}{c}{\boldmath $z=3.00$--$3.50$} & \multicolumn{2}{c}{\boldmath $z=3.50$--$4.25$} \\
\hline
 $\alpha_{\rm sat}$ &   0&983 &   0&981 &   0&975 &   0&986 &   0&991 \\
 $\log_{10}(M_1/M_\odot)$ &  11&443 &  12&236 &  12&346 &  13&277 &  13&322 \\
 $\log_{10}(M_{\star,0}/M_\odot)$ &   9&215 &  10&610 &  10&799 &  11&331 &  11&383 \\
 $\beta$ &   0&307 &   0&353 &   0&364 &   0&534 &   0&530 \\
 $\delta$ &   0&637 &   0&511 &   0&546 &   3&098 &   3&616 \\
 $\gamma$ &   0&565 &   0&998 &   0&795 &   1&592 &   2&682 \\
 $\sigma_{\log M_\star}$ &   0&297 &   0&158 &   0&271 &   0&214 &   0&040 \\
 $B_{\rm cut}$ & 105&434 &  17&484 & 126&670 &  57&653 &  40&166 \\
 $B_{\rm sat}$ &   2&563 &   9&372 &   1&302 &   4&214 &   0&939 \\
 $\beta_{\rm cut}$ &  -0&737 &  -1&044 &  -0&620 &  -4&796 &  -4&318 \\
 $\beta_{\rm sat}$ &   0&402 &   1&389 &   1&496 &   2&345 &   3&031 \\
 $\alpha_\mathrm{sb}$ & \multicolumn{2}{c}{N/A} & \multicolumn{2}{c}{N/A} & \multicolumn{2}{c}{N/A} & \multicolumn{2}{c}{N/A} & \multicolumn{2}{c}{N/A} \\
 $\beta_\mathrm{sb}$ & \multicolumn{2}{c}{N/A} & \multicolumn{2}{c}{N/A} & \multicolumn{2}{c}{N/A} & \multicolumn{2}{c}{N/A} & \multicolumn{2}{c}{N/A} \\
 $\gamma_\mathrm{sb}$ & \multicolumn{2}{c}{N/A} & \multicolumn{2}{c}{N/A} & \multicolumn{2}{c}{N/A} & \multicolumn{2}{c}{N/A} & \multicolumn{2}{c}{N/A} \\
\hline
\end{tabular}

 \begin{tabular}{lr@{.}lr@{.}lr@{.}lr@{.}lr@{.}l}
\hline
 & \multicolumn{2}{c}{\textbf{UKIDSS UDS}} & \multicolumn{2}{c}{\textbf{GAMA}} & \multicolumn{2}{c}{\textbf{PRIMUS}} & \multicolumn{2}{c}{\textbf{PRIMUS}} & \multicolumn{2}{c}{\textbf{PRIMUS}} \\
 \textbf{Parameter} & \multicolumn{2}{c}{\boldmath $z=4.25$--$5.00$} & \multicolumn{2}{c}{\boldmath $z=0.00$--$0.06$} & \multicolumn{2}{c}{\boldmath $z=0.20$--$0.30$} & \multicolumn{2}{c}{\boldmath $z=0.30$--$0.40$} & \multicolumn{2}{c}{\boldmath $z=0.40$--$0.50$} \\
\hline
 $\alpha_{\rm sat}$ &   0&981 &   0&978 &   0&977 &   0&983 &   0&983 \\
 $\log_{10}(M_1/M_\odot)$ &  13&582 &  12&468 &  12&695 &  12&345 &  12&196 \\
 $\log_{10}(M_{\star,0}/M_\odot)$ &  12&141 &  10&749 &  10&922 &  10&775 &  10&593 \\
 $\beta$ &   0&453 &   0&489 &   0&438 &   0&350 &   0&256 \\
 $\delta$ &   2&006 &   0&560 &   1&102 &   0&603 &   0&637 \\
 $\gamma$ &   2&707 &   1&286 &   0&441 &   0&779 &   0&507 \\
 $\sigma_{\log M_\star}$ &   0&011 &   0&117 &   0&077 &   0&013 &   0&178 \\
 $B_{\rm cut}$ &  92&988 &  47&362 &   6&165 &  65&439 &  73&620 \\
 $B_{\rm sat}$ &  15&699 &  11&796 &  11&294 &  14&157 &   7&705 \\
 $\beta_{\rm cut}$ &  -0&048 &  -1&398 &  -2&802 &  -2&046 &  -5&373 \\
 $\beta_{\rm sat}$ &   1&878 &   1&095 &   2&400 &   3&245 &   2&508 \\
 $\alpha_\mathrm{sb}$ & \multicolumn{2}{c}{N/A} &  -1&223 & \multicolumn{2}{c}{N/A} & \multicolumn{2}{c}{N/A} & \multicolumn{2}{c}{N/A} \\
 $\beta_\mathrm{sb}$ & \multicolumn{2}{c}{N/A} &  32&678 & \multicolumn{2}{c}{N/A} & \multicolumn{2}{c}{N/A} & \multicolumn{2}{c}{N/A} \\
 $\gamma_\mathrm{sb}$ & \multicolumn{2}{c}{N/A} &   0&846 & \multicolumn{2}{c}{N/A} & \multicolumn{2}{c}{N/A} & \multicolumn{2}{c}{N/A} \\
\hline
\end{tabular}

 \begin{tabular}{lr@{.}lr@{.}lr@{.}lr@{.}lr@{.}l}
\hline
 & \multicolumn{6}{c}{\textbf{PRIMUS}} & \multicolumn{2}{c}{\textbf{VIPERS}} & \multicolumn{2}{c}{\textbf{VIPERS}} \\
 \textbf{Parameter} & \multicolumn{2}{c}{\boldmath $z=0.50$--$0.65$} & \multicolumn{2}{c}{\boldmath $z=0.65$--$0.80$} & \multicolumn{2}{c}{\boldmath $z=0.80$--$1.00$} & \multicolumn{2}{c}{\boldmath $z=0.50$--$0.60$} & \multicolumn{2}{c}{\boldmath $z=0.60$--$0.80$} \\
\hline
 $\alpha_{\rm sat}$ &   0&979 &   0&984 &   0&989 &   0&981 &   0&982 \\
 $\log_{10}(M_1/M_\odot)$ &  12&362 &  12&188 &  12&817 &  12&938 &  12&510 \\
 $\log_{10}(M_{\star,0}/M_\odot)$ &  10&745 &  10&644 &  10&987 &  10&870 &  10&697 \\
 $\beta$ &   0&288 &   0&026 &   0&655 &   0&488 &   0&299 \\
 $\delta$ &   0&641 &   0&825 &   0&461 &   1&863 &   0&393 \\
 $\gamma$ &   1&162 &   3&648 &   1&755 &   0&681 &   0&924 \\
 $\sigma_{\log M_\star}$ &   0&207 &   0&304 &   0&011 &   0&266 &   0&028 \\
 $B_{\rm cut}$ &  46&790 &  66&882 & 118&198 &  10&945 & 118&137 \\
 $B_{\rm sat}$ &  16&067 &   9&615 &  14&340 &   9&552 &   7&713 \\
 $\beta_{\rm cut}$ &  -2&364 &  -1&065 &  -2&915 &  -2&515 &  -3&339 \\
 $\beta_{\rm sat}$ &   2&060 &   0&278 &   0&140 &   1&666 &   1&236 \\
 $\alpha_\mathrm{sb}$ & \multicolumn{2}{c}{N/A} & \multicolumn{2}{c}{N/A} & \multicolumn{2}{c}{N/A} & \multicolumn{2}{c}{N/A} & \multicolumn{2}{c}{N/A} \\
 $\beta_\mathrm{sb}$ & \multicolumn{2}{c}{N/A} & \multicolumn{2}{c}{N/A} & \multicolumn{2}{c}{N/A} & \multicolumn{2}{c}{N/A} & \multicolumn{2}{c}{N/A} \\
 $\gamma_\mathrm{sb}$ & \multicolumn{2}{c}{N/A} & \multicolumn{2}{c}{N/A} & \multicolumn{2}{c}{N/A} & \multicolumn{2}{c}{N/A} & \multicolumn{2}{c}{N/A} \\
\hline
\end{tabular}

 \end{center}
 \caption{Maximum likelihood parameters of our \protect\HOD\ model for each mass function used as a constraint in this work. Parameters $\alpha_{\rm sat}$ through $\beta_{\rm sat}$ correspond to the \protect\HOD\ model of \protect\cite{behroozi_comprehensive_2010}. Parameters $\alpha_{\rm sb}$, $\beta_{\rm sb}$, and $\sigma_{\rm sb}$ correspond to the surface brightness incompleteness model adopted for the GAMA survey.}
 \label{tb:HODMaximumLikelihood}
\end{table*}

\begin{table*}
 \begin{center}
 \begin{tabular}{lr@{.}lr@{.}lr@{.}lr@{.}lr@{.}l}
\hline
 & \multicolumn{2}{c}{\textbf{VIPERS}} & \multicolumn{2}{c}{\textbf{ZFOURGE}} & \multicolumn{2}{c}{\textbf{ZFOURGE}} & \multicolumn{2}{c}{\textbf{ZFOURGE}} & \multicolumn{2}{c}{\textbf{ZFOURGE}} \\
 \textbf{Parameter} & \multicolumn{2}{c}{\boldmath $z=0.80$--$1.00$} & \multicolumn{2}{c}{\boldmath $z=0.20$--$0.50$} & \multicolumn{2}{c}{\boldmath $z=0.50$--$0.75$} & \multicolumn{2}{c}{\boldmath $z=0.75$--$1.00$} & \multicolumn{2}{c}{\boldmath $z=1.00$--$1.25$} \\
\hline
 $\alpha_{\rm sat}$ &   0&984 &   0&978 &   0&976 &   0&976 &   0&986 \\
 $\log_{10}(M_1/M_\odot)$ &  13&002 &  13&567 &  12&535 &  13&032 &  12&960 \\
 $\log_{10}(M_{\star,0}/M_\odot)$ &  10&846 &  10&984 &  10&580 &  10&987 &  10&916 \\
 $\beta$ &   0&484 &   0&722 &   0&524 &   0&592 &   0&520 \\
 $\delta$ &   3&906 &   1&064 &   0&849 &   1&255 &   2&247 \\
 $\gamma$ &   3&193 &   3&350 &   2&638 &   2&198 &   0&010 \\
 $\sigma_{\log M_\star}$ &   0&271 &   0&551 &   0&390 &   0&192 &   0&225 \\
 $B_{\rm cut}$ & 108&738 &  45&005 &  70&103 & 103&713 &   8&376 \\
 $B_{\rm sat}$ &  14&098 &   0&740 &   5&315 &   1&018 &  11&651 \\
 $\beta_{\rm cut}$ &  -2&907 &  -1&960 &  -1&116 &  -3&693 &  -5&302 \\
 $\beta_{\rm sat}$ &   1&631 &   1&251 &   0&452 &   3&502 &   0&905 \\
 $\alpha_\mathrm{sb}$ & \multicolumn{2}{c}{N/A} & \multicolumn{2}{c}{N/A} & \multicolumn{2}{c}{N/A} & \multicolumn{2}{c}{N/A} & \multicolumn{2}{c}{N/A} \\
 $\beta_\mathrm{sb}$ & \multicolumn{2}{c}{N/A} & \multicolumn{2}{c}{N/A} & \multicolumn{2}{c}{N/A} & \multicolumn{2}{c}{N/A} & \multicolumn{2}{c}{N/A} \\
 $\gamma_\mathrm{sb}$ & \multicolumn{2}{c}{N/A} & \multicolumn{2}{c}{N/A} & \multicolumn{2}{c}{N/A} & \multicolumn{2}{c}{N/A} & \multicolumn{2}{c}{N/A} \\
\hline
\end{tabular}

 \begin{tabular}{lr@{.}lr@{.}lr@{.}lr@{.}lr@{.}l}
\hline
 & \multicolumn{8}{c}{\textbf{ZFOURGE}} & \multicolumn{2}{c}{\textbf{ULTRAVISTA}} \\
 \textbf{Parameter} & \multicolumn{2}{c}{\boldmath $z=1.25$--$1.50$} & \multicolumn{2}{c}{\boldmath $z=1.50$--$2.00$} & \multicolumn{2}{c}{\boldmath $z=2.00$--$2.50$} & \multicolumn{2}{c}{\boldmath $z=2.50$--$3.00$} & \multicolumn{2}{c}{\boldmath $z=0.20$--$0.50$} \\
\hline
 $\alpha_{\rm sat}$ &   0&973 &   0&974 &   0&978 &   0&978 &   0&981 \\
 $\log_{10}(M_1/M_\odot)$ &  13&253 &  12&828 &  12&134 &  13&190 &  12&815 \\
 $\log_{10}(M_{\star,0}/M_\odot)$ &  11&193 &  11&017 &  10&006 &  10&830 &  10&964 \\
 $\beta$ &   0&570 &   0&451 &   0&070 &   0&611 &   0&528 \\
 $\delta$ &   1&231 &   0&691 &   0&341 &   1&533 &   1&019 \\
 $\gamma$ &   0&594 &   0&008 &   1&268 &   3&248 &   0&736 \\
 $\sigma_{\log M_\star}$ &   0&015 &   0&097 &   0&549 &   0&483 &   0&236 \\
 $B_{\rm cut}$ &  28&686 &  90&240 &   4&784 & 122&574 &  72&452 \\
 $B_{\rm sat}$ &   1&237 &   1&662 &  14&508 &   7&255 &  11&449 \\
 $\beta_{\rm cut}$ &  -2&954 &  -2&318 &  -0&257 &  -0&804 &  -5&723 \\
 $\beta_{\rm sat}$ &   2&227 &   1&385 &   3&486 &   0&184 &   3&524 \\
 $\alpha_\mathrm{sb}$ & \multicolumn{2}{c}{N/A} & \multicolumn{2}{c}{N/A} & \multicolumn{2}{c}{N/A} & \multicolumn{2}{c}{N/A} & \multicolumn{2}{c}{N/A} \\
 $\beta_\mathrm{sb}$ & \multicolumn{2}{c}{N/A} & \multicolumn{2}{c}{N/A} & \multicolumn{2}{c}{N/A} & \multicolumn{2}{c}{N/A} & \multicolumn{2}{c}{N/A} \\
 $\gamma_\mathrm{sb}$ & \multicolumn{2}{c}{N/A} & \multicolumn{2}{c}{N/A} & \multicolumn{2}{c}{N/A} & \multicolumn{2}{c}{N/A} & \multicolumn{2}{c}{N/A} \\
\hline
\end{tabular}

 \begin{tabular}{lr@{.}lr@{.}lr@{.}lr@{.}lr@{.}l}
\hline
 & \multicolumn{10}{c}{\textbf{ULTRAVISTA}} \\
 \textbf{Parameter} & \multicolumn{2}{c}{\boldmath $z=0.50$--$1.00$} & \multicolumn{2}{c}{\boldmath $z=1.00$--$1.50$} & \multicolumn{2}{c}{\boldmath $z=1.50$--$2.00$} & \multicolumn{2}{c}{\boldmath $z=2.00$--$2.50$} & \multicolumn{2}{c}{\boldmath $z=2.50$--$3.00$} \\
\hline
 $\alpha_{\rm sat}$ &   0&978 &   0&978 &   0&988 &   0&981 &   0&982 \\
 $\log_{10}(M_1/M_\odot)$ &  12&965 &  12&967 &  13&006 &  12&990 &  13&588 \\
 $\log_{10}(M_{\star,0}/M_\odot)$ &  10&848 &  10&859 &  10&871 &  10&901 &  11&328 \\
 $\beta$ &   0&570 &   0&558 &   0&576 &   0&179 &   0&855 \\
 $\delta$ &   5&027 &   1&809 &   5&529 &   3&658 &   3&567 \\
 $\gamma$ &   4&614 &   3&018 &   4&101 &   3&129 &   1&636 \\
 $\sigma_{\log M_\star}$ &   0&317 &   0&276 &   0&278 &   0&287 &   0&275 \\
 $B_{\rm cut}$ &  10&298 &   7&217 &  93&004 &  51&228 &  40&390 \\
 $B_{\rm sat}$ &  18&991 &  18&652 &  19&213 &   0&944 &   9&906 \\
 $\beta_{\rm cut}$ &  -9&960 &  -5&821 &  -0&876 &  -6&264 &  -0&961 \\
 $\beta_{\rm sat}$ &   2&467 &   2&965 &   0&132 &   3&849 &   0&311 \\
 $\alpha_\mathrm{sb}$ & \multicolumn{2}{c}{N/A} & \multicolumn{2}{c}{N/A} & \multicolumn{2}{c}{N/A} & \multicolumn{2}{c}{N/A} & \multicolumn{2}{c}{N/A} \\
 $\beta_\mathrm{sb}$ & \multicolumn{2}{c}{N/A} & \multicolumn{2}{c}{N/A} & \multicolumn{2}{c}{N/A} & \multicolumn{2}{c}{N/A} & \multicolumn{2}{c}{N/A} \\
 $\gamma_\mathrm{sb}$ & \multicolumn{2}{c}{N/A} & \multicolumn{2}{c}{N/A} & \multicolumn{2}{c}{N/A} & \multicolumn{2}{c}{N/A} & \multicolumn{2}{c}{N/A} \\
\hline
\end{tabular}

 \end{center}
 \addtocounter{table}{-1}
 \caption{\emph{(cont.)}}
\end{table*}

\begin{table}
 \begin{center}
 \begin{tabular}{lr@{.}lr@{.}l}
\hline
 & \multicolumn{2}{c}{\textbf{ULTRAVISTA}} & \multicolumn{2}{c}{\textbf{SDSS (Hearin et al.)}} \\
 \textbf{Parameter} & \multicolumn{2}{c}{\boldmath $z=3.00$--$4.00$} & \multicolumn{2}{c}{\boldmath $z=0.00$--$0.50$} \\
\hline
 $\alpha_{\rm sat}$ &   0&984 &   1&136 \\
 $\log_{10}(M_1/M_\odot)$ &  13&576 &  12&904 \\
 $\log_{10}(M_{\star,0}/M_\odot)$ &  11&469 &  10&775 \\
 $\beta$ &   0&430 &   0&081 \\
 $\delta$ &   3&672 &   0&313 \\
 $\gamma$ &   3&944 &   2&917 \\
 $\sigma_{\log M_\star}$ &   0&435 &   0&396 \\
 $B_{\rm cut}$ &  64&821 & 104&215 \\
 $B_{\rm sat}$ &   8&738 &  22&232 \\
 $\beta_{\rm cut}$ &  -3&098 &  -2&911 \\
 $\beta_{\rm sat}$ &   0&275 &   1&003 \\
 $\alpha_\mathrm{sb}$ & \multicolumn{2}{c}{N/A} & \multicolumn{2}{c}{N/A} \\
 $\beta_\mathrm{sb}$ & \multicolumn{2}{c}{N/A} & \multicolumn{2}{c}{N/A} \\
 $\gamma_\mathrm{sb}$ & \multicolumn{2}{c}{N/A} & \multicolumn{2}{c}{N/A} \\
\hline
\end{tabular}

 \end{center}
 \addtocounter{table}{-1}
 \caption{\emph{(cont.)}}
\end{table}

\begin{figure}
 \includegraphics[width=85mm,trim=0mm 0mm 0mm 0mm,clip]{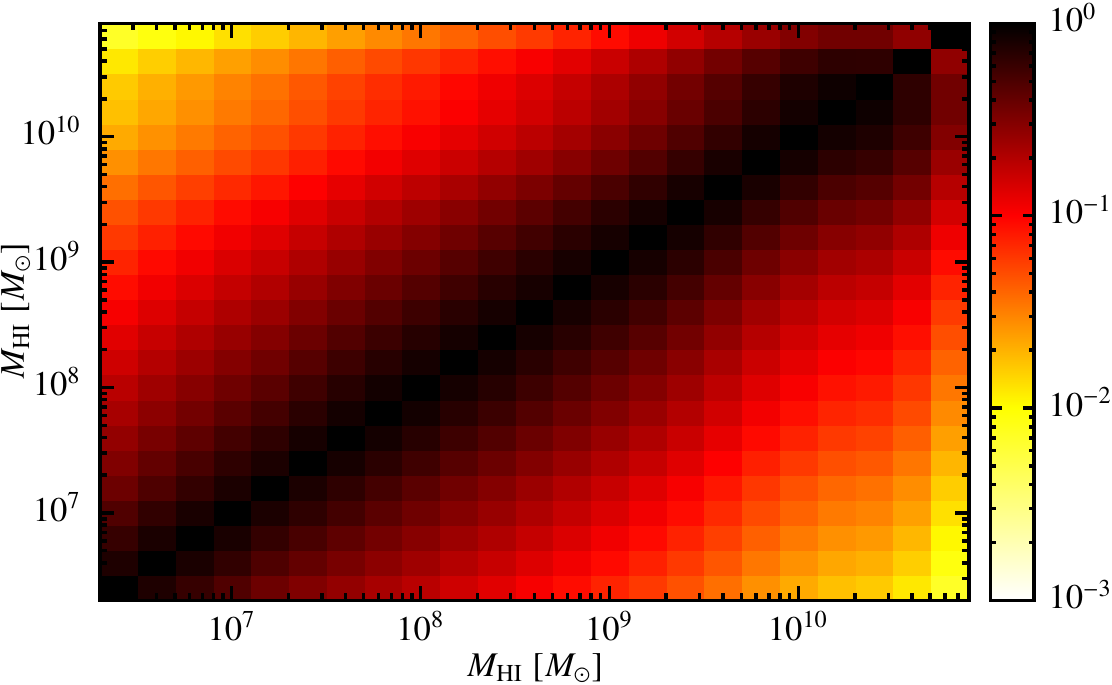}
 \caption{The correlation matrix of the observed galaxy HI mass function of \protect\cite{martin_arecibo_2010}. Colour indicates the strength of correlation between bins, according to the scale shown on the right.}
 \label{fig:CorrelationMatrixALFALFAMain}
\end{figure}

\subsubsection{UKIDSS UDS Stellar Mass Functions}

For stellar mass functions in the interval $z = 3$ to 5 galaxies measured by \cite{caputi_stellar_2011} from the UKIDSS UDS survey, the survey window function is determined from the set of galaxy positions provided by Caputi (private communication), by finding a suitable bounding box and then cutting out empty regions (corresponding to regions that were removed around bright stars). A set of random points are then found within this mask and are used to find the Fourier transform of the survey volume. 

To estimate the depth of the \cite{caputi_stellar_2011} sample as a function of galaxy stellar mass we make use of \SAMs\ in the Millennium Database. We use the \SAMs\ of \cite{guo_dwarf_2011} and \cite{henriques_confronting_2012} ---specifically the {\tt Guo2010a..MR} and {\tt Henriques2012a.wmap1.BC03\_001} tables in the Millennium Database. For each snapshot in the database, we extract the stellar masses and observed-frame IRAC 4.5$\mu$m apparent magnitudes (including dust extinction), and determine the median apparent magnitude as a function of stellar mass. Using the limiting apparent magnitude of the \cite{caputi_stellar_2011} sample, $i_{4.5}=24$, we infer the corresponding absolute magnitude at each redshift and, using our derived apparent magnitude--stellar mass relation, infer the corresponding stellar mass.

\begin{figure}
 \begin{center}
 \includegraphics[width=85mm,trim=0mm 0mm 0mm 4mm,clip]{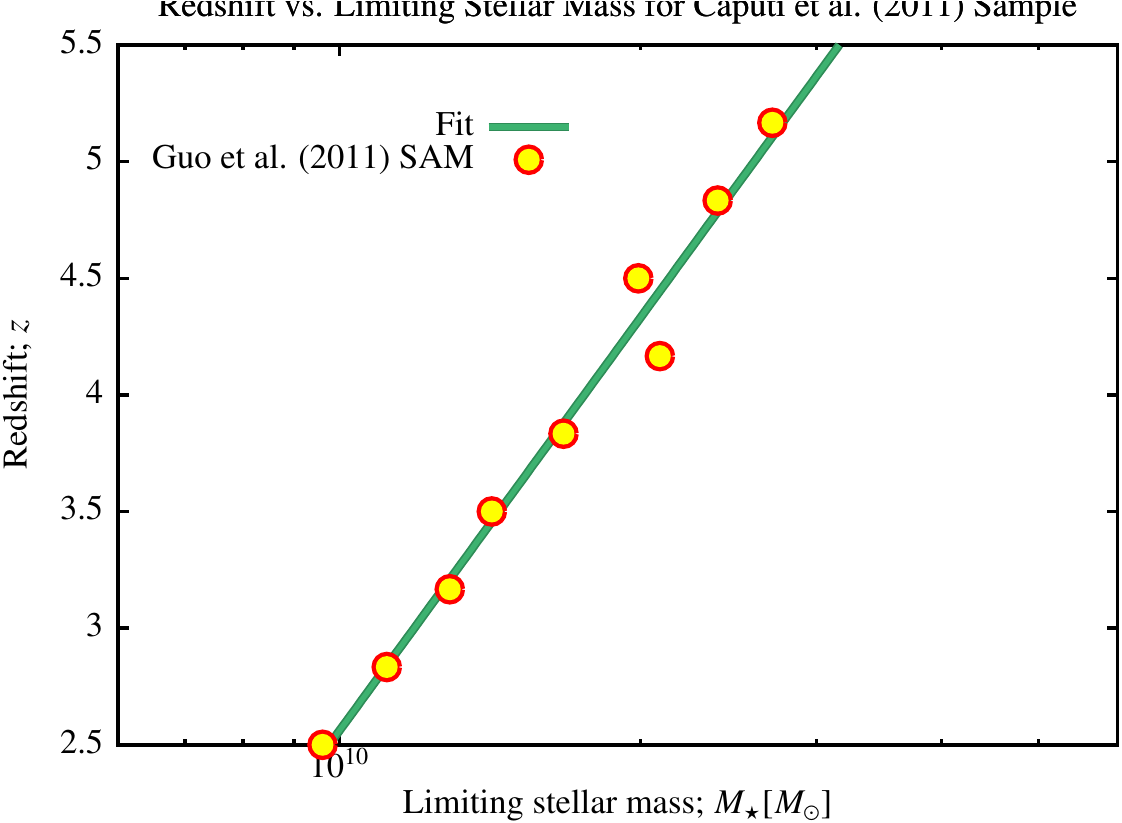}
 \end{center}
 \caption{The maximum redshift at which a galaxy of given stellar mass can be detected in the sample of \protect\cite{caputi_stellar_2011}. Points show the results obtained using the \protect\cite{henriques_confronting_2012} model from the Millennium Database, while the lines shows a polynomial fit to these results (given in eqn.~\ref{eq:UKIDSSUDSDepthPolynomial}).}
 \label{fig:UKIDSSUDSMassRedshift}
\end{figure}

The end result of this procedure is the limiting stellar mass as a function of redshift, accounting for k-corrections, evolution, and the effects of dust. Figure~\ref{fig:UKIDSSUDSMassRedshift} shows the resulting relation between stellar mass and the maximum redshift at which such a galaxy would be included in the sample. Points indicate measurements from the \SAM, while the line shows a polynomial fit:
\begin{equation}
 z(M_\star) = -56.247 + 5.881 m,
 \label{eq:UKIDSSUDSDepthPolynomial}
\end{equation}
where $m= \log_{10}(M_\star/{\rm M}_\odot)$. We use this polynomial fit to determine the depth of the sample as a function of stellar mass.

\begin{figure}
 \begin{center}
 \includegraphics[width=85mm,trim=0mm 0mm 0mm 0mm,clip]{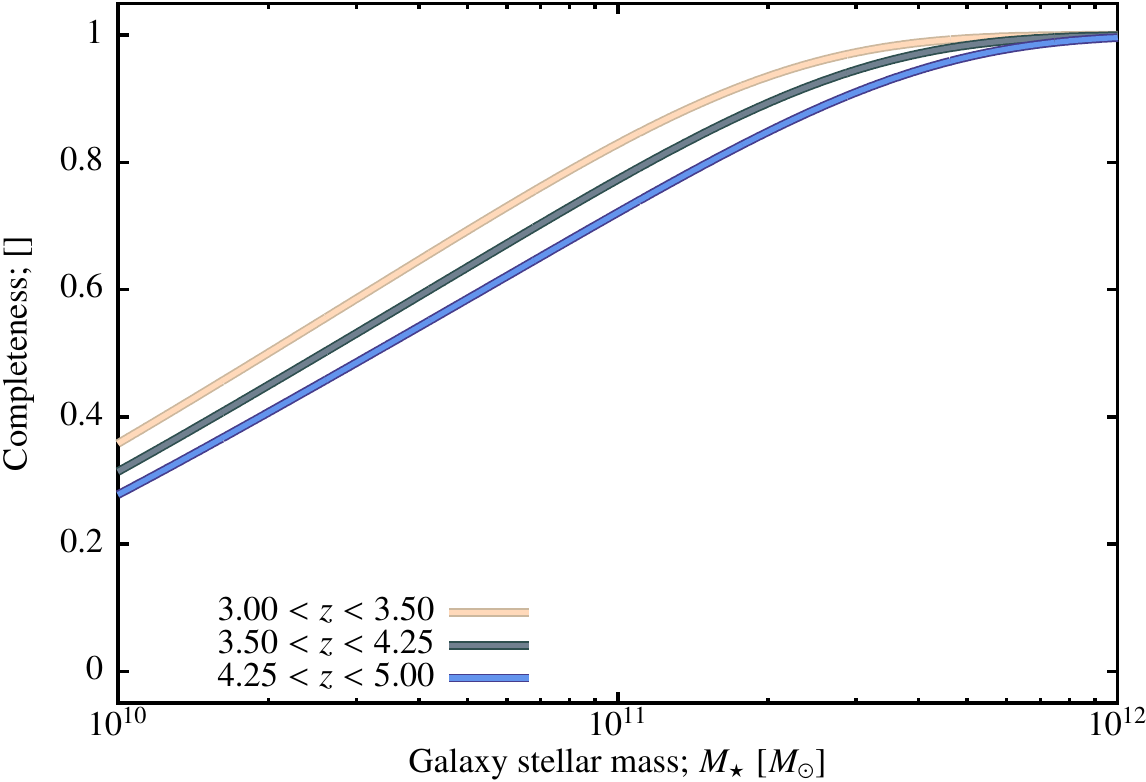}
 \end{center}
 \caption{The completeness as a function of stellar mass in the survey of \protect\cite{caputi_stellar_2011}. Curves are computed using eqn.~(\protect\ref{eq:incompleteness}) with parameters fit to the reported 50\% and 80\% completeness masses from \protect\cite{caputi_stellar_2011}.}
 \label{fig:UKIDSSUDSCompleteness}
\end{figure}

Finally, the incompleteness of the observational sample (which is required when estimating the Poisson contribution to the covariance matrix) is found from the 50\% and 80\% completeness masses, $M_{50}$ and $M_{80}$ respectively, given in Fig.~4 of \cite{caputi_stellar_2011}. Specifically, we assume that, at a given mass $M$, the number of photons assigned to a galaxy can be modeled as a Gaussian distribution with mean $f M$ and variance $fM+\mu$, where $\mu$ is the number of photons arriving from the sky. The fraction of sources of mass $M$ that will be detected at more than $n \sigma$ above the background is then
\begin{eqnarray}
f(M) & = & \int_{n\sqrt{\mu}}^\infty \frac{1}{\sqrt{2 \pi} \sqrt{f M + \mu}} \exp\left( - {[S-fM]^2 \over 2 [f M + \mu]} \right) {\rm d}S \nonumber \\
     & = & {1 \over 2}\left[ 1 - \hbox{erf}\left( {x(M) \over \sqrt{2}} \right)  \right],
\label{eq:incompleteness}
\end{eqnarray}
where $x(M) = (n \sqrt{\mu} - f M)/(\mu+fM)^{1/2}$. Given $f(M_{\rm 50})=0.5$ and $f(M_{\rm 80})=0.8$ we can solve for $f$ and $\mu$, and then compute the completeness in each mass using eqn.~(\ref{eq:incompleteness}). The resulting completeness curves are shown in Fig.~\ref{fig:UKIDSSUDSCompleteness}. Note that the model of eqn.~(\ref{eq:incompleteness}) is clearly an oversimplification, but should capture the expected behavior of the completeness and, since it is fit to the 50\% and 80\% completenesses reported by \cite{caputi_stellar_2011}---which were computed using detailed simulations---should work sufficiently well.

\subsubsection{Li \& White (2009) SDSS Stellar Mass Function}\label{sec:liWhiteSDSS}

For the stellar mass functions at $z \approx 0.07$ galaxies measured by \cite{li_distribution_2009} from the SDSS survey the calculation of the covariance matrix was performed as for \cite{benson_building_2014}, but utilizing the latest version of our code (which contains minor bug fixes relative to the one used in \citealt{benson_building_2014}).

\subsubsection{Bernardi et al. (2013) SDSS Stellar Mass Functions}\label{sec:bernardiSDSS}

To compute the angular mask for the stellar mass functions at $z \approx 0.07$ galaxies measured by \cite{bernardi_massive_2013} from the SDSS survey, we make use of the {\tt mangle} polygon file provided by the {\tt mangle} project\footnote{Specifically, \href{http://space.mit.edu/~molly/mangle/download/data/sdss_dr72safe0_res6d.pol.gz}{http://space.mit.edu/\textasciitilde\ molly/mangle/download/data/sdss\_dr72safe0\_res6d.pol.gz}.} \citep{hamilton_scheme_2004,swanson_methods_2008}. The solid angle of this mask, computed using the {\tt mangle} {\tt harmonize} command is 2.2324~sr.

\begin{figure}
 \begin{center}
 \includegraphics[width=85mm,trim=0mm 0mm 0mm 4mm,clip]{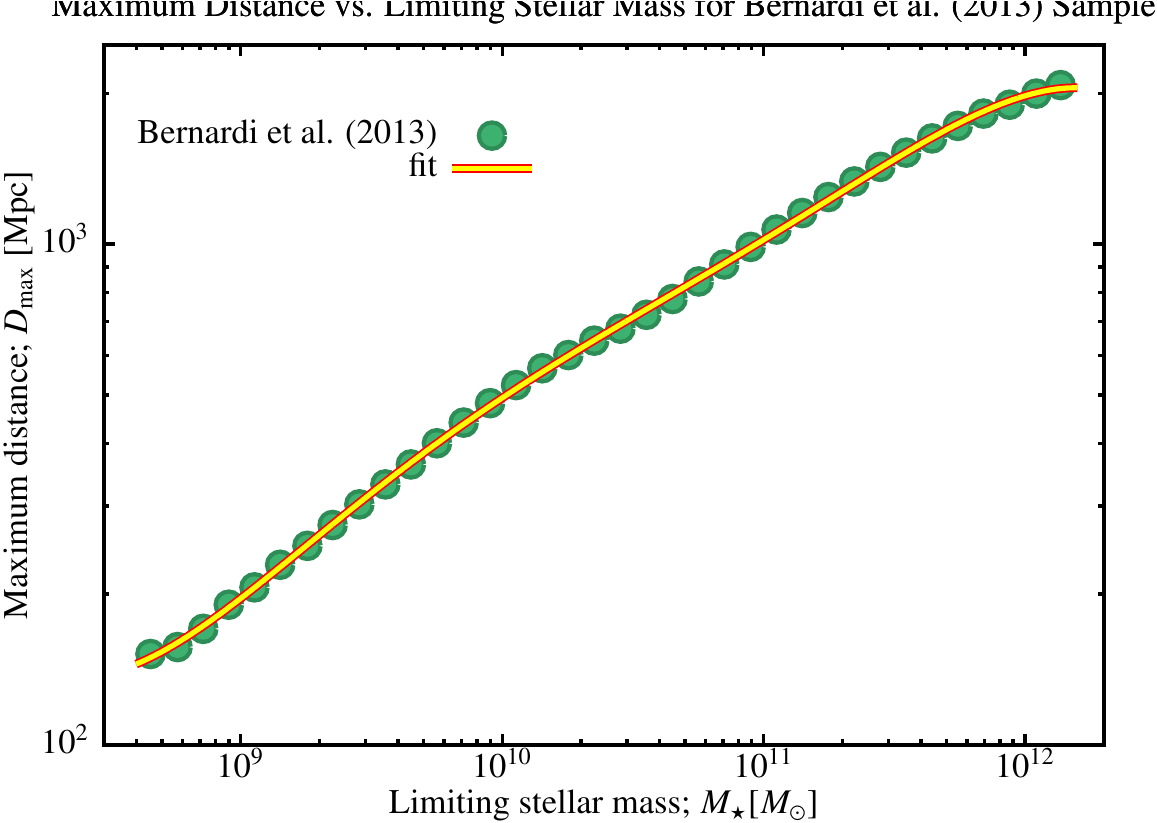}
 \end{center}
 \caption{The maximum distance at which a galaxy of given stellar mass can be detected in the sample of \protect\cite{bernardi_massive_2013}. Points show the results obtained from data provided by Bernardi, while the lines shows a polynomial fit to these results (given in eqn.~\ref{eq:BernardiDepthPolynomial}).}
 \label{fig:BernardiSDSSDepthFit}
\end{figure}

To determine the depth as a function of stellar mass, we make use of results provided by M. Bernardi (private communication), giving the mean maximum volume, $V_{\rm max}$, as a function of stellar mass for galaxies in this sample. These maximum volumes are converted to maximum distances using the solid angle quoted above. The resulting mass vs. distance relation is fit with a $5^{\rm th}$-order polynomial. Figure~\ref{fig:BernardiSDSSDepthFit} shows the resulting relation between stellar mass and the maximum distance at which such a galaxy would be included in the sample. Points indicate results from Bernardi, while the line shows a polynomial fit:
\begin{eqnarray}
 \log_{10} \left[ {D_{\rm max}(M_\star) \over \hbox{Mpc}}\right] = &+& 1282.11 \nonumber \\
&+&m (-626.644\nonumber \\
&+&m (+122.091 \nonumber \\
&+&m (-11.8431\nonumber \\
&+&m (+0.572399\nonumber \\
&+&m (-0.0110301)))))
 \label{eq:BernardiDepthPolynomial}
\end{eqnarray}
where $m= \log_{10}(M_\star/{\rm M}_\odot)$. We use this polynomial fit to determine the depth of the sample as a function of stellar mass.

Finally, the incompleteness of the observational sample is taken to be 91\%\footnote{7\% arising from fiber collisions, 2\% from failures in the Pymorph pipeline (M.~Bernardi, private communication).}.

\subsubsection{Moustakas et al. (2013) PRIMUS Stellar Mass Functions}

To compute the angular mask for stellar mass functions for $z \approx 0.2$ to $z \approx 1.0$ galaxies measured by \cite{moustakas_primus:_2013} from the PRIMUS survey, we make use of {\tt mangle} polygon files provided by J.~Moustakas (private communication)  corresponding to the PRIMUS fields. The solid angle of each mask is computed using the {\tt mangle} {\tt harmonize} command.

\begin{figure}
 \begin{center}
 \includegraphics[width=85mm,trim=0mm 0mm 0mm 3mm,clip]{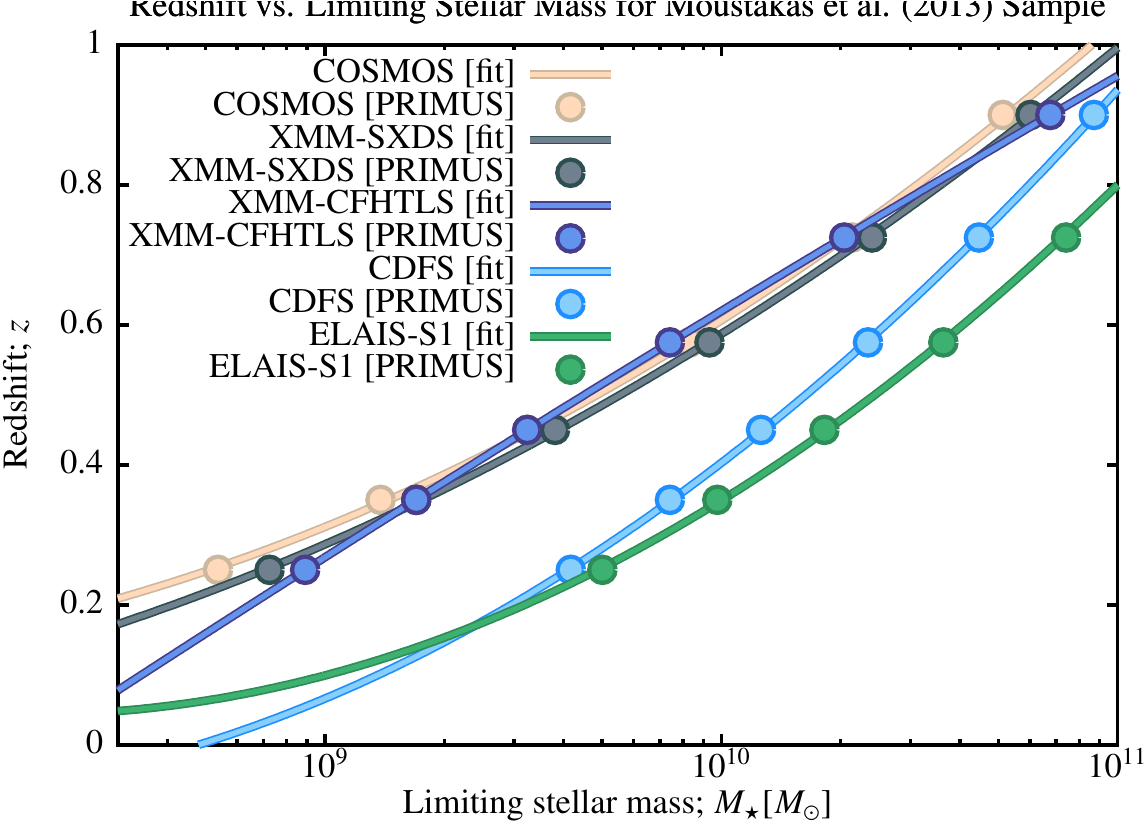}
 \end{center}
 \caption{The maximum distance at which a galaxy of given stellar mass can be detected in the sample of \protect\cite{moustakas_primus:_2013}. Points show the results obtained from completeness limit data taken from Table~2 of \protect\cite{moustakas_primus:_2013}, while the lines shows a polynomial fit to these results (given in eqn.~\ref{eq:MoustakasDepthPolynomial}).}
 \label{fig:MoustakasPRIMUSDepthFit}
\end{figure}

To determine the depth as a function of stellar mass, we make use of completeness limits for ``All'' galaxies given in Table~2 of \cite{moustakas_primus:_2013}. These are fit, for each field, with a second order polynomial to give the limiting redshift as a function of stellar mass. Figure~\ref{fig:MoustakasPRIMUSDepthFit} shows the resulting relation between stellar mass and the maximum redshift at which such a galaxy would be included in the sample. Points indicate results from \cite{moustakas_primus:_2013}, while the lines show polynomial fits:
\begin{equation}
z_{\rm max}(M_\star) = \left\{\begin{array}{ll} +3.51+m(-0.941+m(+0.0651)) & \hbox{\tiny COSMOS} \\ +2.46+m(-0.730+m(+0.0542)) & \hbox{\tiny XMM-SXDS} \\ -3.60+m(+0.500+m(-0.0078)) & \hbox{\tiny XMM-CFHTLS} \\ +5.87+m(-1.528+m(+0.0982)) & \hbox{\tiny CDFS} \\ +6.87+m(-1.656+m(+0.1003)) & \hbox{\tiny ELAIS-S1}\end{array}\right.
 \label{eq:MoustakasDepthPolynomial}
\end{equation}
where $m= \log_{10}(M_\star/{\rm M}_\odot)$. We use this polynomial fit to determine the depth of the sample as a function of stellar mass.

In computing the Poisson contribution to the covariance of the PRIMUS mass function we make use of the actual number of galaxies in each bin, as reported by \cite{moustakas_primus:_2013}. This obviates the need for estimating the completeness in each bin.

\subsubsection{Davidzon et al. (2013) VIPERS Stellar Mass Functions}

To compute angular masks for the stellar mass functions for $z = 0.5$ to $z = 1.0$ galaxies measured by \cite{davidzon_vimos_2013} from the VIPERS survey, we make use of {\tt mangle} polygon files provided by I.~Davidzon (private communication) corresponding to the VIPERS fields. The solid angle of each mask is computed using the {\tt mangle} {\tt harmonize} command.

\begin{figure}
 \begin{center}
 \includegraphics[width=85mm,trim=0mm 0mm 0mm 4mm,clip]{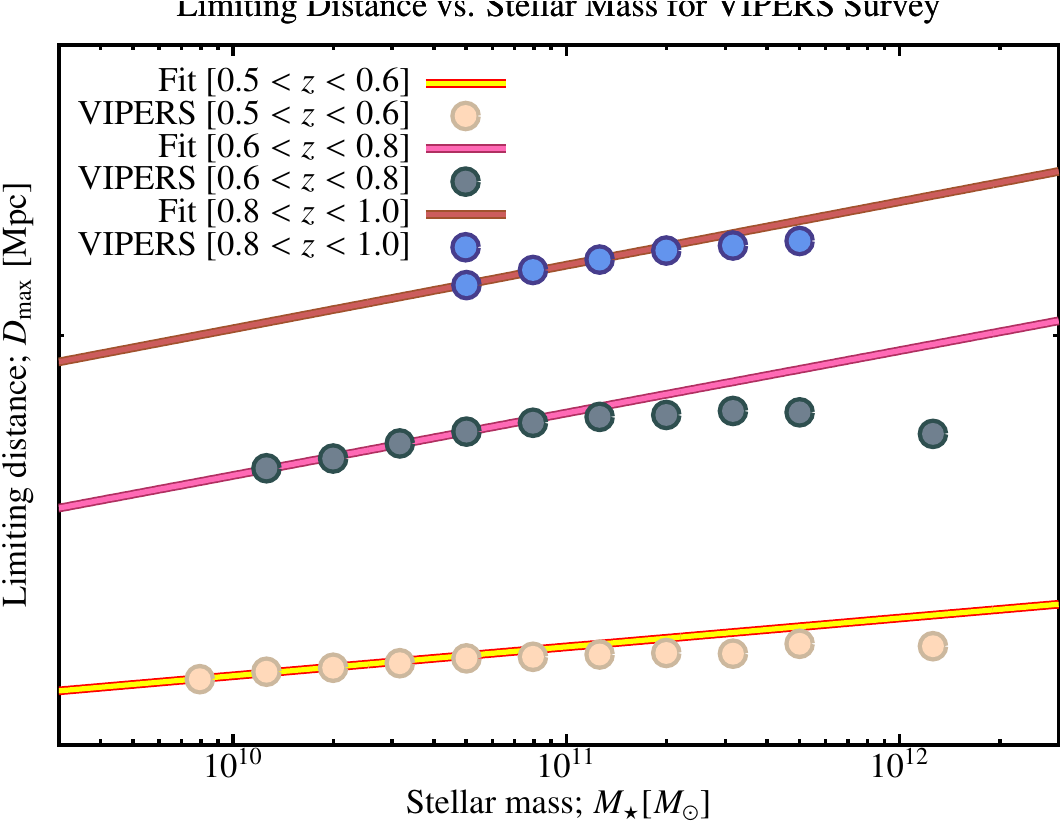}
 \end{center}
 \caption{The maximum distance at which a galaxy of given stellar mass can be detected in the sample of \protect\cite{davidzon_vimos_2013}. Points show the results obtained from data provided by Davidzon, while the lines shows a polynomial fit to these results (given in eqn.~\ref{eq:DavidzonDepthPolynomial}). Note that at high masses the distance is limited by the imposed upper limit---the polynomial fit does not consider these points.}
 \label{fig:Davidzon2013DepthFit}
\end{figure}

To determine the depth as a function of stellar mass, we make use of the tabulated mass function, $\phi$, and number of galaxies per bin, $N$, supplied by I.~Davidzon (private communication\footnote{Note that the mass functions provided were constructed from the same data and using the same techniques as in \protect\cite{davidzon_vimos_2013}, but used different redshift intervals as listed in Table~\protect\ref{tb:HODMaximumLikelihood}.}). The effective volume of each bin is found as $V_i = N_i/f_{\rm complete}\phi_i\Delta\log_{10}M_\star$, where $\Delta\log_{10}M_\star$ is the width of the bin, and $f_{\rm complete}$ is the completeness of the survey, estimated to be approximately 40\% \citep{guzzo_vimos_2013}. These volumes are converted to maximum distances in each field using the field solid angle. The resulting mass vs. distance relation in each field is fit with a $1^{\rm st}$-order polynomial in log-log space over the range where the maximum volume is limited by the survey depth and not by the imposed upper limit to redshift. Figure~\ref{fig:Davidzon2013DepthFit} shows the resulting relation between stellar mass and the maximum distance at which such a galaxy would be included in the sample. Points indicate results from VIPERS, while the lines show polynomial fits:
\begin{equation}
 \log_{10} \left[ {D_{\rm max}(M_\star) \over \hbox{Mpc}}\right] = \left\{ \begin{array}{ll} 3.207 + 0.0124m & 0.5 < z < 0.6 \\ 3.148 + 0.0268m & 0.6 < z < 0.8  \\ 3.207 + 0.0273m & 0.8 < z < 1.0 \end{array} \right.
 \label{eq:DavidzonDepthPolynomial}
\end{equation}
where $m= \log_{10}(M_\star/{\rm M}_\odot)$. We use this polynomial fit to determine the depth of the sample as a function of stellar mass.

\subsubsection{Baldry et al. (2012) GAMA Stellar Mass Functions}\label{sec:GAMAConstraint}

To compute the angular mask for the stellar mass function for $z < 0.06$ galaxies measured by \cite{baldry_galaxy_2012} from the GAMA survey, we use the specifications of the G09, G12, and G15 fields given by \cite{driver_galaxy_2011} to construct {\sc mangle} polygon files from which we compute the survey solid angle and angular power spectrum.

\begin{figure}
 \begin{center}
 \includegraphics[width=85mm,trim=0mm 0mm 0mm 4mm,clip]{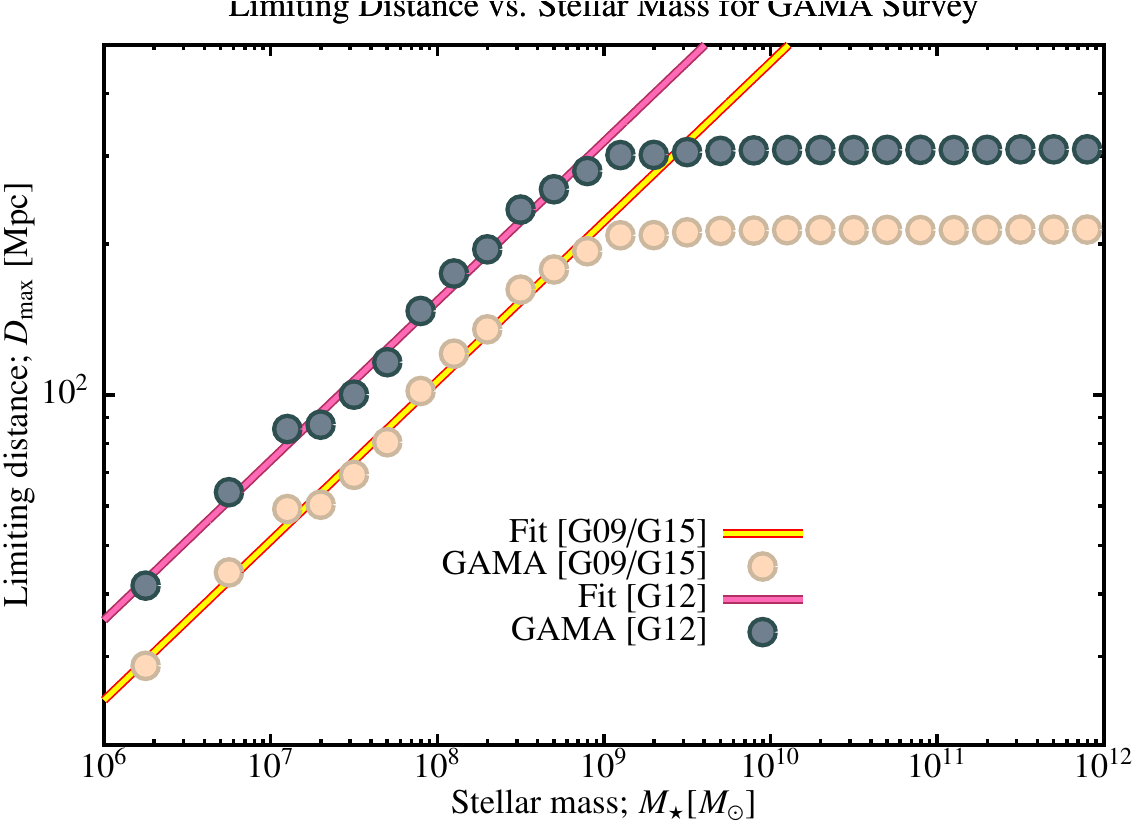}
 \end{center}
 \caption{The maximum distance at which a galaxy of given stellar mass can be detected in the sample of \protect\cite{baldry_galaxy_2012}. Points show the results obtained from data provided by Baldry, while the lines shows a polynomial fit to these results (given in eqn.~\ref{eq:BaldryDepthPolynomial}). Note that above $10^9{\rm M}_\odot$ the distance is limited by the imposed upper limit of $z=0.06$ in the GAMA sample---the polynomial fit does not consider these points.}
 \label{fig:BaldryGAMADepthFit}
\end{figure}

To determine the depth as a function of stellar mass, we make use of the publicly available tabulated mass function, $\phi$, and number of galaxies per bin, $N$. The effective volume of each bin is found as $V_i = N_i/\phi_i\Delta\log_{10}M_\star$, where $\Delta\log_{10}M_\star$ is the width of the bin. The GAMA survey consists of three fields, each of the same solid angle, but with differing depths. We assume that the relative depths in terms of stellar mass scale with the depth in terms of flux. Given this assumption, these volumes are converted to maximum distances in each field using the solid angle quoted above. The resulting mass vs. distance relation in each field is fit with a $1^{\rm st}$-order polynomial in log-log space over the range where the maximum volume is limited by the survey depth and not by the imposed $z=0.06$ upper limit to redshift. Figure~\ref{fig:BaldryGAMADepthFit} shows the resulting relation between stellar mass and the maximum distance at which such a galaxy would be included in the sample. Points indicate results from GAMA, while the line shows a polynomial fit:
\begin{equation}
 \log_{10} \left[ {D_{\rm max}(M_\star) \over \hbox{Mpc}}\right] = \left\{ \begin{array}{ll} -0.521 + 0.319m & \hbox{fields G09/G15} \\ -0.361 + 0.319m & \hbox{field G12} \end{array} \right.
 \label{eq:BaldryDepthPolynomial}
\end{equation}
where $m= \log_{10}(M_\star/{\rm M}_\odot)$. We use this polynomial fit to determine the depth of the sample as a function of stellar mass.

Finally, the completeness of the observational sample is estimated to be greater than 98\% (P.~Norberg, private communication). Therefore we add an additional contribution to the observed covariance matrix equal to $\mathbfss{C}_{ij} = 0.02 \phi_i \phi_j$ where $\phi$ is the observed mass function.

The GAMA mass function is unique within this work in that the reported values extend well into the regime where the survey is incomplete, in this case due to the limiting surface brightness of the survey. As such, \cite{baldry_galaxy_2012} report the values of their mass function in the lowest mass bins as lower limits, as it is clear that a large fraction of galaxies in this mass range are missed. In modeling this mass function we need to construct the true mass function (i.e. without any surface brightness selection effects). We therefore construct a simple model of incompleteness due to surface brightness limits. Specifically, based on Fig.~11 of \cite{baldry_galaxy_2012}, we assume that the distribution of galaxy surface brightness, $\mu$, can be described by a normal distribution with mean
\begin{equation}
 \bar{\mu} = \alpha_\mathrm{sb} \log_{10}(M_\star/{\rm M}_\odot) + \beta_\mathrm{sb},
\end{equation}
where $\alpha_\mathrm{sb}$ and $\beta_\mathrm{sb}$ are parameters, and fixed variance, $\gamma_\mathrm{sb}$. In a given mass bin, the completeness is found by integrating this normal distribution over surface brightnesses brighter than the limiting surface brightness of the GAMA sample, $\mu_{\rm lim}=23.5$~mag~arsec$^{-2}$. Based on Fig.~11 of \cite{baldry_galaxy_2012} we adopt normal priors on $\alpha_\mathrm{sb}$, $\beta_\mathrm{sb}$, and $\gamma_\mathrm{sb}$ with means and variances of $(-1.2,0.0225)$, $(32.7,0.045)$, and $(0.85,0.0025)$ respectively. These parameters are then including in our \MCMC\ analysis when fitting our parametric \HOD\ to the GAMA mass function. The upper limits reported by \cite{baldry_galaxy_2012} can then be treated as actual measurements.

\subsubsection{Tomczak et al. (2014) ZFOURGE Stellar Mass Functions}

To determine the angular mask for stellar mass functions at $z = 0.2$ to $z = 3.0$ galaxies measured by \cite{tomczak_galaxy_2014} from the ZFOURGE survey, we make use of {\tt mangle} polygon files constructed by hand using vertices matched approximately to the distribution of galaxies in the survey (positions of which were provided by R.~Quadri; private communication). The solid angle of each mask is computed using the {\tt mangle} {\tt harmonize} command.

\begin{figure}
 \begin{center}
 \includegraphics[width=85mm,trim=0mm 0mm 0mm 4mm,clip]{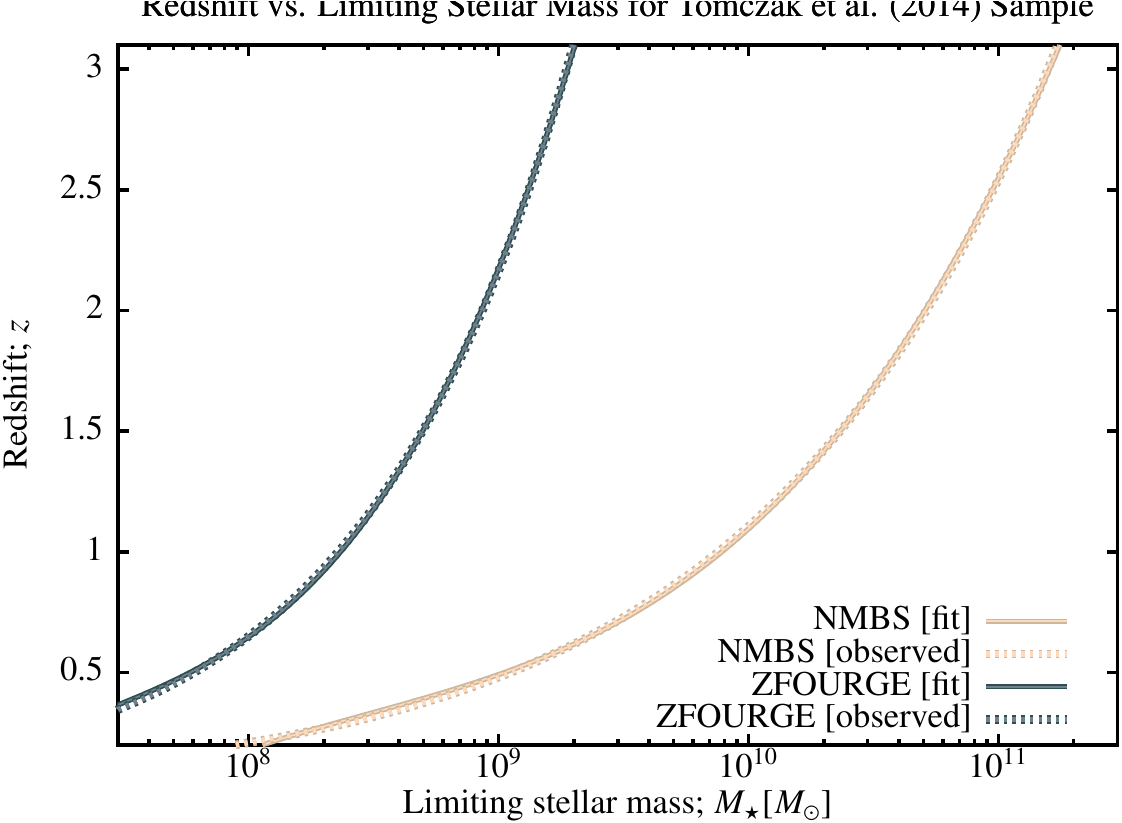}
 \end{center}
 \caption{The maximum redshift at which a galaxy of given stellar mass can be detected in the sample of \protect\cite{tomczak_galaxy_2014}. Points show the results obtained from data provided by Davidzon, while the lines shows a polynomial fit to these results (given in eqn.~\ref{eq:TomczakDepthPolynomial}).}
 \label{fig:Tomczak2014DepthFit}
\end{figure}

To determine the depth as a function of stellar mass, we make use of the tabulated mass completeness limits as a function of redshift for ZFOURGE and NMBS fields provided by R.~Quadri (private communication). These are fit with fourth-order polynomials. Figure~\ref{fig:Tomczak2014DepthFit} shows the resulting relation between stellar mass and the maximum redshift at which such a galaxy would be included in the sample. Dotted lines indicate the tabulated result from ZFOURGE, while the lines show polynomial fits:
\begin{equation}
 z_{\rm max}(M_\star) = \left\{ \begin{array}{ll} -114.66+m(45.901 \\ \,\,+m(-6.1617+m(0.27822))) & \hbox{\tiny ZFOURGE fields} \\ -58.483+m(20.250 \\ \,\,+m(-2.3563+m(0.092705))) & \hbox{\tiny NMBS fields} \end{array} \right.
 \label{eq:TomczakDepthPolynomial}
\end{equation}
where $m= \log_{10}(M_\star/{\rm M}_\odot)$. We use this polynomial fit to determine the depth of the sample as a function of stellar mass.

\subsubsection{Muzzin et al. (2014) ULTRAVISTA Stellar Mass Functions}

To determine the angular mask for stellar mass functions at $z = 0.2$ to $z = 4.0$ measured by \cite{muzzin_evolution_2013} from the ULTRAVISTA survey, we generate a {\sc mangle} polygon file, by first defining a rectangle encompassing the bounds of the ULTAVISTA field ($149.373^\circ < \alpha < 150.779^\circ$ and $1.604^\circ < \delta < 2.81^\circ$). From this rectangle, we then remove circles of radii $75^{\prime\prime}$ around bright stars (i.e. those brighter than 10$^{\rm th}$ and $8^{\rm th}$ magnitudes in the USNO and 2MASS star lists respectively) and radii $30^{\prime\prime}$ around medium stars (i.e. those brighter than $13^{\rm th}$ and $10.5^{\rm th}$ magnitudes in the USNO and 2MASS star lists respectively). Finally, we mask regions of one detector for which 75\% of pixels are dead by clipping pixels with weights below $0.02$ in the K$_{\rm s}$-band weight map. These choices match those made in the ULTRAVISTA survey (A.~Muzzin, private communication). The solid angle of each mask is computed using the {\sc mangle} {\tt harmonize} command.

To determine the depth as a function of stellar mass, we simply fit the tabulated relations\footnote{\url{http://www.strw.leidenuniv.nl/galaxyevolution/ULTRAVISTA/Mstar_redshift_completeness_emp_uvista_v4.1_100.dat}} provided by the ULTRAVISTA survey:
\begin{eqnarray}
z_{\rm max}(M_\star) &=& \left[ -6076.23+m(3231.44+m(-686.816 \right. \nonumber \\
&& \left. +m(72.9148+m(-3.86638+m(0.0819398)))))\right] \nonumber \\
&& \left(1 - \exp[(m-11.24)/0.02]\right)^{-1} 
 \label{eq:MuzzinDepthPolynomial}
\end{eqnarray}
where $m= \log_{10}(M_\star/{\rm M}_\odot)$.

\begin{figure}
 \begin{center}
 \includegraphics[width=85mm,trim=0mm 0mm 0mm 4mm,clip]{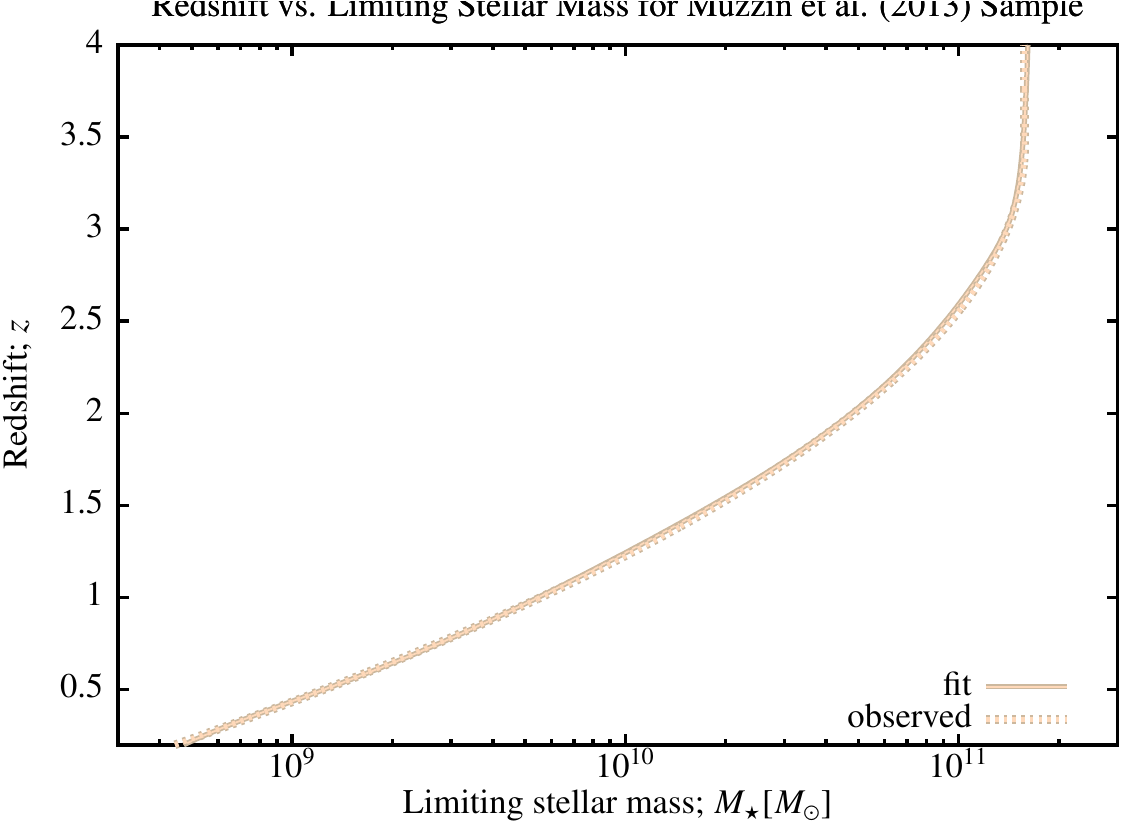}
 \end{center}
 \caption{The maximum redshift at which a galaxy of given stellar mass can be detected in the sample of \protect\cite{muzzin_evolution_2013}. The dotted line shows the results obtained from the ULTRAVISTA survey \protect\citep{muzzin_evolution_2013}, while the solid line shows the polynomial fit to these results (given in eqn.~\ref{eq:MuzzinDepthPolynomial}).}
 \label{fig:Muzzin2014DepthFit}
\end{figure}

\subsubsection{Hearin et al. (2013) SDSS Projected Correlation Functions}

To determine covariances in the projected correlation functions measured by \cite{hearin_dark_2014} in the SDSS we resort to constructing multiple realizations of mock surveys from which we measure the projected correlation function directly. Covariances are then measured from the ensemble of correlation functions. In constructing these mock surveys, the geometry and depth is as described in \S\ref{sec:bernardiSDSS}. We then run an \MCMC\ simulation to constrain the parameters of our \HOD\ model to reproduce the observed projected correlation functions. 

Priors on the parameters of the \HOD\ fit were set to the posterior distribution of our fit to the SDSS stellar mass function (see \S\ref{sec:liWhiteSDSS} and Table~\ref{tb:HODParameterPriors}) such that \HODs\ which match both the mass function and projected correlation functions would be preferred. We find that the posterior distribution is strongly shifted relative to the priors, indicating a strong tension between fitting these projected correlation functions and the SDSS stellar mass function simultaneously---possibly suggesting an insufficiency in our \HOD\ model.

To generate mock survey realizations we first make use of the Bolshoi-P N-body simulation \citep{riebe_multidark_2013}, which has sufficient resolution and volume to permit construction of the survey mocks that we require, and also closely matches the cosmological parameters used in this work. The lowest mass halo populated in our mocks has a mass of approximately $10^{11}{\rm M}_\odot$, corresponding to approximately 500~particles in the Bolshoi-P simulation. As such, all halos used in our mocks are well-resolved and reliable. We extract 25 mock surveys from the Bolshoi-P simulation. For each mock, we select a position uniformly at random within the box, and choose a random line of sight from an isotropic distribution to correspond to the center of the SDSS survey field. We then select all halos which lie within (or close to) the volume defined by the SDSS angular mask and depth. We populate each such halo with a number of central and satellite galaxies drawn at random from the \HOD, with centrals placed at the halo centre, and satellites tracing a \cite{navarro_universal_1997} profile centred on the halo centre and with scale radius selected from the concentration-mass relation of \cite{gao_redshift_2008}---this avoids relying on poorly measured concentrations in low particle number N-body halos. The projected correlation function is then measured directly from each mock in the same bins\footnote{Both separations, $r_\mathrm{p}$, and the projected correlation function, $w_\mathrm{p}$, in \protect\cite{hearin_dark_2014} were reported in units of Mpc. However, the numerical values listed in that work were actually in units of Mpc$/h$ (A. Hearin, private communication). We have therefore corrected for this factor of $h$ when constructing the projected correlation function in this work.} as used by \cite{hearin_dark_2014} using a \cite{landy_bias_1993} estimator (with random points generated using the same angular mask and depth used to build the mock catalogs).

Given the limited number of independent survey volumes that can be extracted from the Bolshoi-P simulation we also generate mock surveys using the Pinocchio algorithm \citep{monaco_pinocchio_2002,monaco_pinocchio:_2013}. We generate 100 independent Pinocchio simulations with a box size of 290~Mpc, using a $1024^3$ grid. This size represents a compromise between resolution and volume---ideally a larger volume would be used to avoid the need to replicate the simulation cube when constructing mock catalogs. Halo catalogs are output at $z=0.05$ and are used to construct mock galaxy surveys in the same way as was used for the Bolshoi-P simulation.

\begin{figure}
 \includegraphics[width=85mm,trim=0mm 0mm 0mm 0mm,clip]{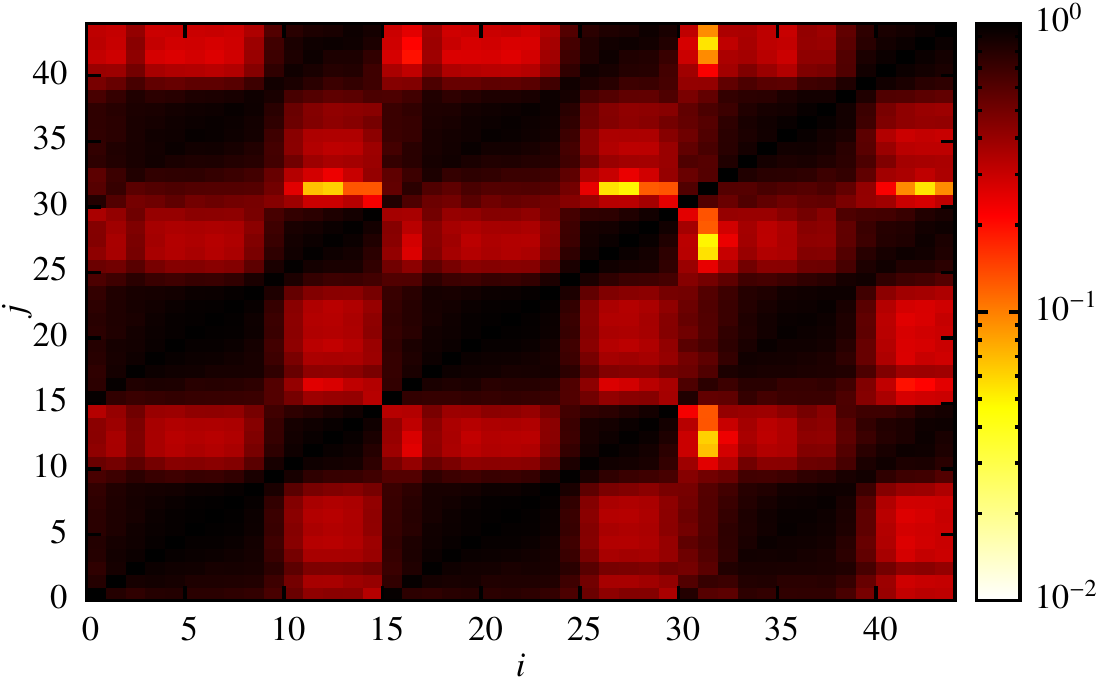} 
 \caption{The correlation matrices of the observed galaxy projected correlation functions of \protect\cite{hearin_dark_2014}. Colour indicates the strength of correlation between bins, according to the scale shown on the right. The $3\times 3$ block nature of the correlation matrix reflects the three mass ranges used by \protect\cite{hearin_dark_2014}.}
 \label{fig:CorrelationMatrixSDSSClustering}
\end{figure}

Covariance matrices estimated from simulations are noisy and biased \citep{joachimi_errors_2014}. To mitigate this problem we use the shrinkage technique of \cite{pope_shrinkage_2008}. Specifically, we take the covariance matrix measured from the Bolshoi-P simulation as our empirical estimate of the covariance matrix and that measured from the Pinocchio mocks as our target (since it has smaller variance but may be biased as Pinocchio does not precisely reproduce the statistics of N-body simulation halos). The resulting correlation matrix is shown in Figure~\ref{fig:CorrelationMatrixSDSSClustering}. This is a $3\times 3$ block matrix with each block representing one of the three mass ranges for which the projected correlation function was measured by \cite{hearin_dark_2014}. Along the block diagonals (i.e. looking at the covariance of the projected correlation function within a single mass range) we find that there is very strong correlation between bins in the projected correlation function---in particular it is noticeable that all small separation points are strongly correlated with each other, as are all large separation points (with the break occurring at around $r_{\rm p}\approx 4$~Mpc (where clustering is transitioning between linear and non-linear). This reflects the regimes in which the one- and two-halo terms dominate in the halo model of clustering. Figure~\ref{fig:CorrelationMatrixSDSSClustering} also shows that there is substantial correlation between points in projected correlation functions corresponding to different mass samples. This occurs both because lower mass samples include all galaxies from the higher mass samples, and because the halos involved all sample the same large scale structure. Clearly, accounting for covariance in correlation function measurements is very important when they are used to constrain models.

\section{Discussion}

We have computed estimates of the covariance matrices for galaxy stellar mass functions as reported by several different authors, and for one measurement of the projected correlation function. Our approach is based on modeling the observed mass and correlation functions using an \HOD\ approach, coupled with either halo model analytical estimates of the covariance (for the mass functions; \citealt{smith_how_2012}), or Monte Carlo estimation using approximate simulations (for the projected correlation function).

In all cases we find significant covariance between measured data points, arising due to the presence of large scale structure within the survey volume, and due to the intrinsically correlated nature of galaxies (which are frequently found in groups and clusters). As many theoretical models now make use of observational datasets of this kind to derive quantitative constraints on model parameters, it is important that the likelihood functions upon which that inference is built be quantitatively accurate. Incorporating the covariance in observational measurements is an important component of that goal \citep{benson_building_2014}.

The approach described here relies on an \HOD\ approach. In its current form this means that the effects of environment and assembly bias are ignored (i.e. galaxy properties are assumed to depend only upon the mass of the halo in which the galaxy lives). The approach could straightforwardly be generalized to account for correlations of galaxy properties if a suitable parameterized model were developed (see, for example \citealt{hearin_dark_2014}). Also in the current approach we treat each dataset separately. In principle a single model could be developed which simultaneously fits all of the observations of interest (see, for example, \citealt{behroozi_average_2013}), and then used to construct covariance matrices for all datasets. This would have the advantage of providing a better constrained model in cases where an individual observational dataset is itself not very constraining. Models which could potentially be used in this way include UniverseMachine \citep{behroozi_universemachine:_2018}, EMERGE \citep{moster_emerge_2018}, EMERGE, and various \SAMs\ \citep[see]{baugh_primer_2006,benson_galaxy_2010}. Using such models to compute covariances would also allow application of more complex observational selection effects, and for the computation of covariances between different datasets.

All of the covariance matrices described in this work are made freely available at\ldots\footnote{Files will be made available once this paper is accepted for publication.}

\section*{Acknowledgments}

We thank Martha Haynes for helpful discussion on the ALFALFA data set, John Moustakas for providing masks and data from the PRIMUS survey, Peder Norberg and Ivan Baldry for providing data from the GAMA survey, Mariangela Bernardi for providing mass function data, Iary Davidzon for providing masks and data from the VIPERS survey, Ryan Quadri and Adam Muzzin for providing data and masks from the ZFOURGE survey, Andrew Hearin and Andreas Berlind for providing covariance matrices for galaxy correlation functions, Andrew Hamilton, Molly Swanson, and Max Tegmark for discussions on their {\sc mangle} software (developed also by Colin Hill), Warren Perger for providing a code to compute the generalized hypergeometric function, the members of the Astrostatistics Facebook group for invaluable discussions, and Richard Bower for encouraging me to finish this work.

Funding for the Sloan Digital Sky Survey (SDSS) has been provided by the Alfred P. Sloan Foundation, the Participating Institutions, the National Aeronautics and Space Administration, the National Science Foundation, the U.S. Department of Energy, the Japanese Monbukagakusho, and the Max Planck Society. The SDSS Web site is \href{http://www.sdss.org/}{\tt http://www.sdss.org/}.

The SDSS is managed by the Astrophysical Research Consortium (ARC) for the Participating Institutions. The Participating Institutions are The University of Chicago, Fermilab, the Institute for Advanced Study, the Japan Participation Group, The Johns Hopkins University, Los Alamos National Laboratory, the Max-Planck-Institute for Astronomy (MPIA), the Max-Planck-Institute for Astrophysics (MPA), New Mexico State University, University of Pittsburgh, Princeton University, the United States Naval Observatory, and the University of Washington. 

The CosmoSim database used in this paper is a service by the Leibniz-Institute for Astrophysics Potsdam (AIP). The Bolshoi simulations have been performed within the Bolshoi project of the University of California High-Performance AstroComputing Center (UC-HiPACC) and were run at the NASA Ames Research Center.

This research has made use of NASA's Astrophysics Data System (ADS) and the arXiv preprint server.

We acknowledge the Simons Foundation for supporting the series of ``Galactic Superwinds'' symposia, which helped motivate much of the work in this paper, and the hospitality of the Kavli Institute for Theoretical Physics where part of this work was completed. This research was supported in part by the National Science Foundation under Grant No. NSF PHY11-25915.

Computing resources used in this work were made possible by a grant from the Ahmanson Foundation. We made extensive use of several open source software projects, including GCC, Python, Perl, HDF5, GSL, FGSL, FoX, and FFTW.

\bibliographystyle{mn2e}
\bibliography{covariancesAccented}

\appendix

\section{Covariance Matrix File Format}\label{sec:fileFormat}

The covariance matrices computed in this work are made available as HDF5 files. For mass functions, each file contains the following datasets\footnote{Note that all masses and mass functions are reported under a consistent set of units and definitions as described in the text. In most cases this has required conversion of the values reported by the original authors to account for their choice of logarithm base, inclusion of ``$h$'' factors, etc.}:
\begin{description}
\item[\texttt{mass}] the masses ($M$, in units of $\mathrm{M}_\odot$) corresponding to the mid-point of each bin in which the mass function was measured;
\item[\texttt{massFunction}] the observed mass function ($\mathrm{d}n/\mathrm{d}\log_\mathrm{e} M$, in units of $\mathrm{Mpc}^{-3}$) as reported by the original authors;
\item[\texttt{covariance}] the total covariance matrix of the \texttt{massFunction} dataset ($\mathbfss{C}$, in units of $\mathrm{Mpc}^{-6}$) as computed in this work;
\item[\texttt{covariancePoisson}] the contribution of the Poisson term to the covariance matrix ($\mathbfss{C}_\mathrm{Poisson}$, in units of $\mathrm{Mpc}^{-6}$);
\item[\texttt{covarianceLSS}] the contribution of the large scale structure term to the covariance matrix ($\mathbfss{C}_\mathrm{LSS}$, in units of $\mathrm{Mpc}^{-6}$);
\item[\texttt{covarianceHalo}] the contribution of the halo term to the covariance matrix ($\mathbfss{C}_\mathrm{halo}$, in units of $\mathrm{Mpc}^{-6}$);
\item[\texttt{correlation}] the correlation matrix ($\mathbfss{R}=\mathbfss{D}^{-1} \mathbfss{C} \mathbfss{D}^{-1}$, where $\mathbfss{D} = \sqrt{\mathrm{diag}(\mathbfss{C})}$);
\item[\texttt{inverseCovariance}] the inverse of the covariance matrix ($\mathbfss{C}^{-1}$);
\item[\texttt{logDeterminantCovariance}] the (natural) logarithm of the determinant of the covariance matrix ($\log_\mathrm{e} |\mathbfss{C}|$).
\end{description}
The latter two datasets are included as a convenience for computation of log-likelihoods.

For the projected correlation function, the file contains the following datasets:
\begin{description}
\item[\texttt{massMinimum}] the minimum stellar mass used in selecting galaxies for each of the three samples ($M_{\star,\mathrm{min}}$, in units of $\mathrm{M}_\odot$);
\item[\texttt{massMaximum}] the maximum stellar mass used in selecting galaxies for each of the three samples ($M_{\star,\mathrm{max}}$, in units of $\mathrm{M}_\odot$);
\item[\texttt{separation}] the projected separation at the centre of each bin in which the correlation function is measured ($r_\mathrm{p}$, in units of Mpc);
\item[\texttt{projectedCorrelationFunction}] the projected correlation function in each bin ($w(r_\mathrm{p})$, in units of Mpc; this dataset contains the correlation function for all three mass samples);
\item[\texttt{covariance}] the total covariance matrix of the \texttt{projectedCorrelationFunction} dataset ($\mathbfss{C}$, in units of $\mathrm{Mpc}^2$) as computed in this work (this is a $3\times 3$ block matrix, reflecting the three mass samples used by \citealt{hearin_dark_2014});
\item[\texttt{correlation}] the correlation matrix ($\mathbfss{R}=\mathbfss{D}^{-1} \mathbfss{C} \mathbfss{D}^{-1}$, where $\mathbfss{D} = \sqrt{\mathrm{diag}(\mathbfss{C})}$);
\item[\texttt{inverseCovariance}] the inverse of the covariance matrix ($\mathbfss{C}^{-1}$);
\item[\texttt{logDeterminantCovariance}] the (natural) logarithm of the determinant of the covariance matrix ($\log_\mathrm{e} |\mathbfss{C}|$).
\end{description}


Datasets containing dimensionful quantities have additional attributes as follows:
\begin{description}
\item[\texttt{units}] the units of the dataset in human-readable form;
\item[\texttt{unitsInSI}] the factor by which the dataset should be multiplied to convert to SI units.
\end{description}

Each file also contains several attributes which store relevant quantities which were used in the calculation of the covariance matrix\footnote{The calculations were performed using the {\sc Galacticus} toolkit \protect\citep{benson_galacticus:_2012}.}:
\begin{description}
\item[\texttt{OmegaMatter}] the matter density parameter, $\Omega_\mathrm{M}$;
\item[\texttt{OmegaDarkEnergy}] the dark energy density parameter, $\Omega_\Lambda$;
\item[\texttt{OmegaBaryon}] the baryon density parameter, $\Omega_\mathrm{M}$;
\item[\texttt{HubbleConstant}] the Hubble parameter, $H_0$, in units of km/s/Mpc;
\item[\texttt{hodAlphaSatellite}] parameter of the \cite{leauthaud_new_2012} \HOD\ model, $\alpha_\mathrm{sat}$;
\item[\texttt{hodBetaCut}]  parameter of the \cite{leauthaud_new_2012} \HOD\ model, $\beta_\mathrm{cut}$;
\item[\texttt{hodBetaSatellite}] parameter of the \cite{leauthaud_new_2012} \HOD\ model, $\beta_\mathrm{sat}$;
\item[\texttt{hodBCut}] parameter of the \cite{leauthaud_new_2012} \HOD\ model, $B_\mathrm{cut}$;
\item[\texttt{hodBSatellite}] parameter of the \cite{leauthaud_new_2012} \HOD\ model, $B_\mathrm{sat}$;
\item[\texttt{hodBeta}] parameter of the \cite{behroozi_comprehensive_2010} \SHMR\ model, $\beta$;
\item[\texttt{hodDelta}] parameter of the \cite{behroozi_comprehensive_2010} \SHMR\ model, $\delta$;
\item[\texttt{hodGamma}] parameter of the \cite{behroozi_comprehensive_2010} \SHMR\ model, $\gamma$;
\item[\texttt{hodLog10M1}] parameter of the \cite{behroozi_comprehensive_2010} \SHMR\ model, $\log_{10}M_1$;
\item[\texttt{hodLog10Mstar0}] parameter of the \cite{behroozi_comprehensive_2010} \SHMR\ model, $\log_{10}M_{\star,0}$;
\item[\texttt{hodSigmaLogMstar}] parameter of the \cite{leauthaud_new_2012} \HOD\ model, $\sigma_{\log_{10} M_\star}$;
\end{description}

\onecolumn

\section{Large Scale Structure Covariance Term Using Spherical Harmonics}\label{sec:lssCovariance}

In evaluating the large scale structure contribution to mass function covariance we must evaluate the variance
\begin{equation}
 \sigma^2(M_\mu,M_\nu) = \int \frac{{\rm d}^3 {\bf k}}{(2 \pi)^3} P(k) W(k|M_\mu) W^*(k|M_\nu),
\end{equation}
where $P(k)$ is the nonlinear matter power spectrum (typically averaged over the redshifts over which the mass function is measured), and $W(k|M)$ is the Fourier transform of the window function associated with the survey volume for galaxies of mass $M$:
\begin{equation}
 W(k|M_\mu) = \frac{1}{V_\mu} \int {\rm d}^3{\bf x} \exp(i{\bf k}\cdot{\bf x})\Phi({\bf x}|M_\mu),
\end{equation}
where $\Phi({\bf x}|M)$ is the survey window function for galaxies of mass $M$. If our survey consists of multiple fields, possibly each with different depths, then this window function can be written as a sum over the product of angular and radial parts of each field such that
\begin{equation}
 W(k|M_\mu) = {1\over V_\mu} \int {\rm d}^3{\bf x} \exp(i{\bf k}\cdot{\bf x}) \sum_i \psi^i({\bf \Omega}) \rho^i(r|M_\mu),
\end{equation}
where $i$ runs over fields, and where we explicitly assume that the angular component, $\psi({\bf \Omega})$, is independent of mass. We wish to express this in terms of the spherical harmonic coefficients of the angular mask. We begin by expanding the plane wave in the above in terms of spherical harmonics, giving us
\begin{equation}
 W(k|M_\mu) = {4 \pi \over V_\mu} \sum_i \sum_{\ell=0}^\infty i^{-\ell} \int {\rm d}r\, r^2 j_\ell(kr) \rho^i(r|M_\mu) \sum_{m=-\ell}^{+\ell} Y_{\ell m}(\theta^\prime,\phi^\prime) \int {\rm d}{\bf \Omega} Y^*_{\ell m}(\theta,\phi) \psi^i({\bf \Omega}),
\end{equation}
where $(\theta^\prime,\phi^\prime)$ defines the direction of the $k$-vector, and $j_\ell(x)$ is the spherical Bessel function. The final integral is just the usual expression for the coefficients of the spherical harmonics expansion of $\psi^i({\bf \Omega})$, so
\begin{equation}
 W(k|M_\mu) = {4 \pi \over V_\mu} \sum_i \sum_{\ell=0}^\infty i^{-\ell} \int {\rm d}r\, r^2 j_\ell(kr) \rho^i(r|M_\mu) \sum_{m=-\ell}^{+\ell} Y_{\ell m}(\theta^\prime,\phi^\prime) \Psi^i_{\ell m}.
\end{equation}
If we approximate the radial part of the window function as equal to $1$ within the radial range of the survey, and $0$ outside of that range then, defining
\begin{equation}
 R_{\ell}(x_0,x_1) \equiv \int_{x_0}^{x_1} x^2 j_\ell(x) {\rm d}x = \sqrt{\pi} 2^{-2-\ell} \Gamma\left({1\over 2}[3+\ell]\right) \left[ x^{3+\ell} \tensor*[_1]{\stackrel{\sim}{F}}{_2} \left({1\over 2}[3+\ell]; \ell+{3\over 2},{1\over 2}(5+\ell);-{x^2\over 4}\right)\right]_{x_0}^{x_1},
\end{equation}
where $\tensor*[_1]{\stackrel{\sim}{F}}{_2}$ is the regularized generalized hypergeometric function, we find
\begin{equation}
 W(k|M_\mu) = {4 \pi \over k^3 V_\mu} \sum_i \sum_{\ell=0}^\infty i^{-\ell} R_{\ell}(kr^i_{\mu 0},kr^i_{\mu 1}) \sum_{m=-\ell}^{+\ell} Y_{\ell m}(\theta^\prime,\phi^\prime) \Psi^i_{\ell m}.
\end{equation}
Our expression for the variance now becomes
\begin{eqnarray}
 \sigma^2(M_\mu,M_\nu) &=& {2 \over \pi V_\mu V_\nu} \sum_i \sum_j \int {\rm d}^3 {\bf k} {P(k) \over k^6} \sum_{\ell=0}^\infty i^{-\ell} R^i_{\ell}(kr_{\mu 0},kr_{\mu 1}) \sum_{\ell^\prime=0}^\infty i^{-\ell^\prime} R^j_{{\ell^\prime}}(kr_{\nu 0},kr_{\nu 1}) \sum_{m=-\ell}^{+\ell} Y_{\ell m}(\theta^\prime,\phi^\prime) \Psi^i_{\ell m} \nonumber \\
 & & \times \sum_{m^\prime=-\ell^\prime}^{+\ell^\prime} Y^*_{\ell^\prime m^\prime}(\theta^\prime,\phi^\prime) \Psi^{j*}_{\ell^\prime m^\prime}.
\end{eqnarray}
Using the orthonormality of the spherical harmonics this reduces to:
\begin{equation}
  \sigma^2(M_\mu,M_\nu) = {2 \over \pi V_\mu V_\nu}\int_0^\infty {\rm d} k\, k^{-4} P(k) \sum_i \sum_j \sum_{\ell=0}^\infty (2\ell+1) C^{ij}_\ell R^i_{\ell}(kr_{\mu 0},kr_{\mu 1}) R^j_{\ell}(kr_{\nu 0},kr_{\nu 1}),
\end{equation}
where $(2\ell+1) C^{ij}_\ell = \sum_{m=-\ell}^{+\ell} \Psi^i_{\ell m} \Psi^{j*}_{\ell m}$.

\end{document}